\documentclass[%
twocolumn,
superscriptaddress,
 amsmath,amssymb,
 aps,prb,
]{revtex4-2}

\usepackage{graphicx}
\usepackage{dcolumn}
\usepackage{bm}
\usepackage{color}
\usepackage{ulem}
\usepackage{xspace}
\usepackage{multirow}
\usepackage[unicode=true,colorlinks=true,linkcolor=blue,citecolor=blue,urlcolor=blue]{hyperref}
\usepackage{physics}
\usepackage{braket}
\usepackage{here}
\usepackage[utf8]{inputenc}

\usepackage{comment}
\usepackage{newtxtext}

\definecolor{green}{rgb}{0,0.6,0.1}

\begin{document}

\preprint{APS/123-QED}

\title{Anharmonic Gr\"uneisen theory based on self-consistent phonon theory: Impact of phonon-phonon interaction neglected in the quasiharmonic theory}
\author{Ryota Masuki}
\email{masuki-ryota774@g.ecc.u-tokyo.ac.jp}
\affiliation{
Department of Applied Physics, The University of Tokyo,7-3-1 Hongo, Bunkyo-ku, Tokyo 113-8656
}

\author{Takuya Nomoto}
\affiliation{
Department of Applied Physics, The University of Tokyo,7-3-1 Hongo, Bunkyo-ku, Tokyo 113-8656
}

\author{Ryotaro Arita}
\affiliation{
Department of Applied Physics, The University of Tokyo,7-3-1 Hongo, Bunkyo-ku, Tokyo 113-8656
}
\affiliation{ 
RIKEN Center for Emergent Matter Science, 2-1 Hirosawa, Wako, Saitama 351-0198, Japan 
}
\author{Terumasa Tadano}
\affiliation{ 
CMSM, National Institute for Materials Science (NIMS), 1-2-1 Sengen, Tsukuba, Ibaraki 305-0047, Japan
}


\date{\today}

\begin{abstract}
We formulate a theory of thermal expansion based on the self-consistent phonon (SCP) theory, which nonperturbatively considers the anharmonic effect. We show that the Gr\"unseisen formula holds within the SCP theory by replacing the phonon frequency by the SCP frequency.
By comparing it with the quasiharmonic approximation (QHA), we derive explicit formulae of the correction to the QHA result.
We show that the phonon anharmonicity gives a small correction of $O(\braket{\hat{U}_4}/\braket{\hat{U}_2})$
to the thermal expansion coefficient $\alpha$, where $\hat{U}_2$ and $\hat{U}_4$ are the harmonic and the quartic terms of the potential energy surface. 
On the other hand, we show that the phonon anharmonicity gives two correction terms to the temperature ($T$)-dependent phonon frequency shift which are comparable to the original QHA term.
In strongly anharmonic materials such as NaCl and MgO, these two correction terms tend to cancel out each other, which explains why QHA sometimes gives reasonable values for the $T$-dependent phonon frequency shift while it fails for thermal expansion.
\end{abstract}

\maketitle


\section{Introduction} 
Controlling thermal expansion is one of the most crucial goals in material science because the volume change of materials is often problematic in various situations.
Recently, materials that show negative thermal expansion have been intensively studied for their potential application to cancel out the total thermal expansion of objects~\cite{Mary90, doi:10.1021/cm9602959, doi:10.1021/ja056460f, doi:10.1021/ja106711v, PhysRevLett.121.255901,Dove_2016}.
Therefore, understanding the physics of thermal expansion and clarifying the application limit of the approximations in the calculation are essential for the effective search for materials with desired properties, as well as interesting as a problem of fundamental science.

The quasiharmonic approximation (QHA) is the most widely used approximation in the first-principles calculation of thermal expansion of materials since it gives reasonable results for the thermal expansion coefficient $\alpha = \frac{1}{V} (\partial V/\partial T)_P$ of many materials with relatively small computational cost~\cite{PhysRevB.61.8793, PhysRevB.71.205214, doi:10.1063/1.4879543, PhysRevB.92.081408,Ritz_JAP2019}.
QHA is based on the assumption that the potential energy surface (PES) is nearly harmonic and disregards all the anharmonic effects except for the volume($V$)-dependence of the phonon frequency.
However,
the effect of the neglected intrinsic anharmonicity, i.e., phonon-phonon interaction, on thermal expansion is still unclear, which needs to be clarified for understanding the physics of thermal expansion and rationalizing the assumption of QHA~\cite{PhysRevB.92.064106, doi:10.1142/S0217984920500256}.

In this research, we start from the microscopic anharmonic Hamiltonian and formulate a theory for thermal expansion based on the self-consistent phonon (SCP) theory~\cite{doi:10.1080/14786435808243224, PhysRevLett.17.89, PhysRevB.1.572, doi:10.7566/JPSJ.87.041015, PhysRevB.92.054301, PhysRevMaterials.3.033601,  PhysRevB.89.064302, PhysRevB.96.014111}, which incorporates the effect of the lattice anharmonicity in a nonperturbative way.
While SCP theory has been employed for the numerical calculation of stress tensor, pressure, and thermal expansion of materials\cite{PhysRev.165.951, PhysRevB.98.024106}, the effect of phonon anharmonicity to the thermal expansion and related phenomena has not been fully clarified by analytical calculation.
We prove that the Gr\"uneisen formula exactly holds within the SCP theory by redefining the Gr\"uneisen parameter using the SCP phonon frequency. 
This result also applies to strongly anharmonic materials for which QHA breaks down.
We compare the theory with the quasiharmonic theory and derive an explicit form of the correction that the lattice anharmonicity gives to the physical quantities.
We show that the lattice anharmonicity gives a correction of 
$O(\braket{\hat{U}_4}/\braket{\hat{U}_2})$
to the thermal expansion coefficient $\alpha$, where $\hat{U}_2$ and $\hat{U}_4$ are the harmonic and the quartic terms of the PES.
On the other hand, we show that the anharmonicity gives 
a correction that consists of two 
different contributions to the temperature($T$)-dependent phonon frequency shift,
which are both comparable to the QHA term.

Our theory explains why QHA often fails to correctly describe the 
$T$-dependent phonon frequency shift even for harmonic materials. For example, the optical mode of silicon softens around seven times more rapidly than the prediction of QHA when the temperature rises. In addition, QHA gives the wrong sign for the $T$-dependent shift of the transverse acoustic (TA) mode of silicon~\cite{PhysRevB.91.014307, Kim1992, doi:10.1063/1.5125779}.
It has seemed like a paradox that QHA quantitatively reproduces thermal expansion although QHA fails to accurately calculate the phonon frequency shift because the thermal expansion coefficient $\alpha$ can be rewritten by using the Gr\"uneisen parameter
\begin{equation}
    \gamma^{\text{QHA}}_{\bm{k}\lambda}(V) = - \frac{V}{\omega_{\bm{k}\lambda}} \frac{d \omega_{\bm{k}\lambda}}{d V},
\end{equation}
according to the Gr\"uneisen theory~\cite{doi:10.1063/1.5125779,https://doi.org/10.1002/andp.19123441202}, which is based on the same assumption as the QHA. 
Here, $\omega_{\bm{k}\lambda}(V)$ is the volume ($V$) dependent frequency of the $\lambda$-th phonon with wave number ${\bm k}$.
This Gr\"uneisen parameter $\gamma_{\bm{k}\lambda}$ describes the phonon frequency shift when the system undergoes thermal expansion.
Our theory validates that QHA successfully applies to the thermal expansion of weakly anharmonic materials that satisfy 
$|\braket{\hat{U}_4}/\braket{\hat{U}_2}| \ll 1$
, even when QHA fails to correctly reproduce the $T$-dependent phonon frequency shift.

Furthermore, we perform the first-principles calculations on 
several insulators such as silicon and diamond (covalent crystals), NaCl (ionic crystal), and MgO (oxide) to numerically test our theory and quantify the applicable limit of QHA.
Notably, the QHA works better for the phonon frequency shift than for the thermal expansion in NaCl and MgO, which have strong anharmonicity. 
This is because the two terms of the anharmonic correction to the phonon frequency shift have opposite signs when $\braket{\hat{U}_4} > 0$, and they cancel out each other. We speculate that this fortuitous cancellation occurs in a broad range of anharmonic materials. On the other hand, 
our results make it clear that the agreement of the phonon frequency shift with the experimental result
does not theoretically justify the use of QHA to materials with strong anharmonicity.

\section{Theory}
\label{sec_theory}

In this paper, we consider the isotropic case, in which the expansion of the system is parametrized by a single parameter $V$ (the system volume) or $a$ (the lattice constant).
It is straightforward to extend the discussion below to the anisotropic cases.
The formulae we derive in this section are summarized in Table \ref{table_summary_formulae}.
\subsection{The Gr\"uneisen theory of thermal expansion and the quasiharmonic approximation (QHA)}
\label{subsec_the_Gruneisen_theory}
We first review the Gr\"uneisen theory of thermal expansion~\cite{doi:10.1063/1.5125779,https://doi.org/10.1002/andp.19123441202} and the quasiharmonic approximation (QHA), which are equivalent theories.
See Appendix \ref{Appendix_sec_Gruneisen_formula} for the details of the derivation.

In the Gr\"uneisen theory and QHA, it is assumed
that the PES is nearly harmonic at each volume, i.e., we disregard all the lattice anharmonicity except for the volume dependence of the phonon frequency $\omega_{\bm{k}\lambda}(V)$. Then, the corresponding free energy is 
\begin{multline}
F_{\text{QHA}}(V,T) \\
= E_{\text{gnd}}(V) + \sum_{\bm{k}\lambda} \Bigl[ \frac{1}{2} \hbar \omega_{\bm{k}\lambda}(V) + k_{\mathrm{B}} T \log (1- e^{-\beta \hbar \omega_{\bm{k}\lambda}(V)}) \Bigr],
\label{eq_QHA_free_energy}
\end{multline}
where $E_{\text{gnd}}(V)$ is the $V$-dependent electron ground state energy, and $k_{B}$ is the Boltzmann constant.

In the QHA, the $T$-dependence of the volume is determined by minimizing $F_{\text{QHA}}(V, T)$ at each temperature as 
$V(T) = \arg \mathrm{min}_{V} F_{\text{QHA}}(V, T) $
. The $F_{\text{QHA}}$ is calculated by using the $E_{\text{gnd}}(V)$ and $\omega_{\bm{k}\lambda}(V)$, which can be obtained using the first-principles calculation.

In the Gr\"uneisen theory, we continue the analytical calculation to get the Gr\"uneisen formula
\begin{align}
\alpha^{\text{QHA}} = \frac{1}{B^{\text{QHA}}_T} \frac{1}{V}\sum_{\bm{k}\lambda} c^{\text{QHA}}_{v,\bm{k}\lambda} \gamma^{\text{QHA}}_{\bm{k}\lambda},
\label{eq:alpha_gruneisen_QHA}
\end{align}
where $B_T$ is the bulk modulus. $ c^{\text{QHA}}_{v,\bm{k}\lambda}$ is the mode specific heat 
\begin{align}
 c^{\text{QHA}}_{v,\bm{k}\lambda}(V,T) = \hbar  \omega_{\bm{k}\lambda} \frac{\partial  n_B(\hbar \omega_{\bm{k}\lambda} )}{\partial T},
\end{align}
with $n_B(\hbar \omega)$ being the Bose--Einstein distribution function,
and $\gamma^{\text{QHA}}_{\bm{k}\lambda}$ is the Gr\"uneisen parameter
\begin{align}
\gamma^{\text{QHA}}_{\bm{k}\lambda} (V) = - \frac{V}{\omega_{\bm{k}\lambda}(V)} \Bigl( \frac{d \omega_{\bm{k}\lambda}(V)}{d V} \Bigr).
\end{align}
From the assumption of the theory, the $T$-dependent phonon frequency shift is
\begin{equation}
  \frac{d \omega_{\bm{k}\lambda}}{d T} 
  = \frac{d \omega_{\bm{k}\lambda}(V)}{d V} \frac{d V}{d T} 
 =
  - \alpha^{\text{QHA}} \omega_{\bm{k}\lambda} \gamma^{\text{QHA}}_{\bm{k}\lambda}.
  \label{eq_Gruneisen_freq_shift}
\end{equation}

\subsection{The self-consistent phonon (SCP) theory}
The self-consistent phonon (SCP) theory is based on the variational principle of the free energy. 
The \textit{effective} harmonic Hamiltonian
\begin{equation}
\hat{\mathcal{H}}_0 = \sum_{\bm{k}\lambda} \hbar \Omega_{\bm{k}\lambda} \Bigl(\hat{n}_{\bm{k}\lambda} + \frac{1}{2}\Bigr),
\end{equation}
is employed as the trial Hamiltonian,
where the frequencies $\Omega_{\bm{k}\lambda}$ are the variational parameters.
We assume that the change of the polarization vectors by the anharmonic renormalization can be neglected and use the fixed-mode approximation.
The variational free energy is 
\begin{align}
&&
\mathcal{F}_1 (V, T, \Omega)
  = 
  \mathcal{F}_0 + \braket{\hat{H} - \hat{\mathcal{H}}_0}_{\hat{\mathcal{H}}_0},
\end{align}
where $\mathcal{F}_0 = - k_{\mathrm{B}} T \log \Tr e^{-\beta \mathcal{H}_0}$.
By calculating the stationary condition of the variational free energy with respect to the variational parameters, 
we get the SCP equation
\begin{align}
&\Omega_{\bm{k}\lambda}^2 =
\omega_{\bm{k}\lambda}^2 
  \nonumber
  +
  \sum_{n=2}^\infty \frac{1}{(n-1)! N^{n-1}} \sum_{\bm{k}_1\lambda_1, \cdots, \bm{k}_{n-1}\lambda_{n-1}} 
  \nonumber
  \\&
  \times
  \widetilde{\Phi}(\bm{k} \lambda, -\bm{k} \lambda, \bm{k}_1 \lambda_1, -\bm{k}_1 \lambda_1, \cdots, \bm{k}_{n-1} \lambda_{n-1}, -\bm{k}_{n-1} \lambda_{n-1})
  \nonumber
  \\&\times 
  g(\Omega_{\bm{k}_1\lambda_1}) \cdots g(\Omega_{\bm{k}_{n-1}\lambda_{n-1}}),
  \label{eq_SCP_equation}
\end{align}
where $\widetilde{\Phi}(\bm{k} \lambda, -\bm{k} \lambda, \cdots, \bm{k}_{n-1} \lambda_{n-1}, -\bm{k}_{n-1} \lambda_{n-1})$ is the reciprocal space representation of the $2n$-th order interatomic force constant (IFC), which is defined in Appendix \ref{Appendix_sec_Taylor_expansion}.
Here, we define 
$g(\Omega) = \frac{\hbar}{2\Omega}( n_B(\hbar \Omega) + \frac{1}{2} )$
for notational simpliticy.
For the detailed derivation of the SCP equation, see Appendix \ref{Appendix_sec_SCP}.
We call the $\Omega_{\bm{k}\lambda}$, which is the solution of the SCP equation, as the SCP frequency.
This SCP equation, which is based on the Taylor expansion of the PES, is a generalization of the previous results~\cite{PhysRevB.92.054301} to infinite orders and is equivalent to the other forms of the SCP equation~\cite{PhysRevLett.17.89, PhysRevB.89.064302}.

However, this SCP frequency is not necessarily interpreted as the experimentally-observed $T$-dependent phonon frequency.
Instead, one should use the Hessian of the SCP free energy
~\cite{PhysRevB.96.014111}.
Here, we consider the diagonal part of the Hessian of the SCP free energy because we fix the polarization vector. 
When a static atomic displacement is introduced in the system, the expectation values of the normal coordinate operators $\hat{q}_{\bm{k}\lambda}$, which is defined in Appendix \ref{Appendix_sec_Taylor_expansion}, become finite; we denote this as $q_{\bm{k}\lambda}$, without the hat on $q$.
We calculate the Hessian for the uniform displacement ($\bm{k} = 0$) in Appendix \ref{Appendix_sec_SCP}, which gives
\begin{align}
  \frac{\partial^2 \mathcal{F}_1}{\partial q^{2}_{\bm{0}\lambda}}
  \nonumber\
  \simeq&
  \Omega_{\bm{0}\lambda}^2
  \nonumber
  \\&+
  \sum_{\bm{k}\lambda_1} \frac{\hbar}{4} \frac{|\widetilde{\Phi}(-\bm{k}\lambda_1, \bm{k}\lambda_1, \bm{0}\lambda)|^2}{\Omega_{\bm{k}\lambda_1^2}} 
  \nonumber
  \\&\times
  \Bigl( \frac{\partial n_B(\hbar \Omega_{\bm{k}\lambda_1})}{\Omega_{\bm{k}\lambda_1}}  - \frac{2n_B(\hbar \Omega_{\bm{k}\lambda_1}) + 1}{2\Omega_{\bm{k}\lambda_1}}\Bigr)
  \nonumber
  \\&+
  \sum_{\bm{k}\lambda_1 \lambda_2(\Omega_{\bm{k}\lambda_1} \neq \Omega_{\bm{k}\lambda_2} )} 
  \frac{\hbar}{4} \frac{| \widetilde{\Phi}(-\bm{k}\lambda_1, \bm{k}\lambda_2, \bm{0}\lambda) |^2}{ \Omega_{\bm{k}\lambda_1} \Omega_{\bm{k}\lambda_2} }
  \nonumber
  \\&\times
  \Bigl(
    \frac{n_B(\hbar \Omega_{\bm{k}\lambda_1}) - n_B(\hbar \Omega_{\bm{k}\lambda_2})}{\Omega_{\bm{k}\lambda_1} - \Omega_{\bm{k}\lambda_2}}
    \nonumber
    \\&
    - 
    \frac{n_B(\hbar \Omega_{\bm{k}\lambda_1}) + n_B(\hbar \Omega_{\bm{k}\lambda_2})+1}{\Omega_{\bm{k}\lambda_1} + \Omega_{\bm{k}\lambda_2}}
  \Bigr)
  \label{eq_QP0_squared_freq}
\end{align}
This result reproduces the $\bm{k} = 0$ case of the SCP+QP[0] theory~\cite{tadano2021firstprinciples}
implemented in the ALAMODE package,
which adds the static ($\omega = 0$) part of the bubble self-energy to the quasi particle (QP) phonon frequencies, 
 and is consistent with the previous result that uses another formalism of the SCP theory~\cite{PhysRevB.96.014111}.
While the SCP free energy incorporates only the even-order anharmonicity, this Hessian frequency includes the odd-order anharmonicity as well.
We can derive the more general result of the Hessian for finite $\bm{k}$ in a similar way by considering a commensurate supercell.
When the Hessian is positive-semidefinite, we define the frequency calculated from this Hessian as
\begin{align}
    \Omega^{\text{Hes}}_{\bm{k}\lambda} = \sqrt{\frac{\partial^2 \mathcal{F}_1}{\partial q_{\bm{k}\lambda} \partial q^{*}_{\bm{k}\lambda} } },
    \label{eq_def_Hessian_freq}
\end{align}
where the right-hand side (RHS) can be calculated similarly as Eq. (\ref{eq_QP0_squared_freq}).
For later convenience, we define the frequency shift from the SCP frequency  as
\begin{align}
    \Delta \Omega^{\text{Hes}}_{\bm{k}\lambda} = \Omega^{\text{Hes}}_{\bm{k}\lambda} - \Omega_{\bm{k}\lambda}
\end{align}
It should be noted again that we interpret this $\Omega^{\text{Hes}}_{\bm{k}\lambda}$, not the SCP frequency $\Omega_{\bm{k}\lambda}$, as the frequency measured by the experiments such as the inelastic neutron scattering.

\subsection{SCP theory of thermal expansion}
In the SCP theory of thermal expansion, the optimum volume is calculated by minimizing the SCP free energy, as in QHA. In this subsection, we derive an analytical formula for the thermal expansion coefficient based on the SCP theory, in analogy with the Gr\"uneisen theory.

We start with the SCP entropy, which is calculated by differentiating the SCP free energy $\mathcal{F}_1$ and organizing the result using the SCP equation,
\begin{align}
S(V,T)
  =
-\sum_{\bm{k}\lambda} \bigl[ k_{\mathrm{B}} \log (1 - e^{-\beta\hbar \Omega_{\bm{k}\lambda}}) - \frac{\hbar \Omega_{\bm{k}\lambda}}{T} n_B(\hbar \Omega_{\bm{k}\lambda})\bigr],
\end{align}
which is consistent with the results derived from the different formalisms of SCP~\cite{PhysRevB.1.572, Hui_1975}.
See Appendix \ref{Appendix_sec_SCP_entropy} for the detailed derivation of the SCP entropy.
The SCP entropy has the same form as the entropy of the harmonic Hamiltonian except that the phonon frequency is replaced by $\Omega_{\bm{k}\lambda}$. Starting from this SCP entropy, we follow the scheme of the derivation of the Gr\"uneisen theory, which is explained in detail in Appendix \ref{Appendix_sec_Gruneisen_formula}. Then, we prove that the Gr\"uneisen formula [Eq.~(\ref{eq:alpha_gruneisen_QHA})] rigorously holds for the SCP theory of the thermal expansion
\begin{equation}
\alpha = \frac{1}{B_T} \frac{1}{V}\sum_{\bm{k}\lambda} c_{v,\bm{k}\lambda} \gamma_{\bm{k}\lambda},
\label{eq:alpha_gruneisen_SCP}
\end{equation}
which have been overlooked in the previous research.
Note that the mode specific heat $ c_{v,\bm{k}\lambda}$ and the Gruneisen parameter $\gamma_{\bm{k}\lambda}$ are redefined using the SCP frequency $\Omega_{\bm{k}\lambda}$ as 
\begin{align}
 c_{v,\bm{k}\lambda}(V,T) &= \hbar  \Omega_{\bm{k}\lambda}  \Bigl(\frac{\partial  n_B(\hbar \Omega_{\bm{k}\lambda} )}{\partial T} \Bigr)_{\Omega}
 \nonumber
 \\&=
 \frac{(\hbar \Omega_{\bm{k}\lambda})^2}{k_{\mathrm{B}}T^2} n_B(\hbar \Omega_{\bm{k}\lambda}) (n_B(\hbar \Omega_{\bm{k}\lambda}) + 1),
\end{align}
\begin{align}
\gamma_{\bm{k}\lambda}(V,T) = - \frac{V}{\Omega_{\bm{k}\lambda}(V,T)} \Bigl( \frac{\partial \Omega_{\bm{k}\lambda}(V,T)}{\partial V} \Bigr).
\end{align}
It should be noted that $ c_{v,\bm{k}\lambda}$ here is not the mode specific heat in a rigorous sense because $\frac{1}{N_k}\sum_{\bm{k}\lambda}c_{v,\bm{k}\lambda} \neq C_{V}$ where $C_V$ is the total heat capacity of SCP. However, we conventionally call $c_{v,\bm{k}\lambda}$ the mode specific heat. This result applies to the strongly anharmonic cases.
The $T$-dependent shift of the anharmonic phonon frequency is
\begin{align}
  \Bigl(\frac{\partial \Omega_{\bm{k}\lambda}^{\text{Hes}}}{\partial T} \Bigr)_{P} = - \alpha \Omega_{\bm{k}\lambda} \gamma_{\bm{k}\lambda} + \left( \frac{\partial \Omega_{\bm{k}\lambda}}{\partial T} \right)_{V} + \left( \frac{\partial \Delta \Omega^{\text{Hes}}_{\bm{k}\lambda}}{\partial T} \right)_{P},
  \label{eq_anharmonic_freq_shift}
\end{align}
where additional two term appear compared to Eq. (\ref{eq_Gruneisen_freq_shift}).
This gives us insight that the QHA is a good approximation for calculating the thermal expansion but not for the frequency shift. 
In the following subsection, we perform perturbation expansion to examine this idea.

\subsection{The perturbation expansion}
\label{sec_perturbation_expansion}
We consider the case in which we can treat the lattice anharmonicity as a perturbation and investigate how the lattice anharmonicity affects the thermal expansion and the $T$-dependent phonon frequency shift.

In this subsection, we implicitly assume that the system volume $V$ at which we calculate $\omega_{\bm{k}\lambda}$ and $\Omega_{\bm{k}\lambda}$ are the same when comparing QHA and SCP. However, this is not exactly true because the former, which we call the QHA volume $V_{\text{QHA}}$, is determined by minimizing the $F_{\text{QHA}}$ and the latter, which we call the SCP volume $V_{\text{SCP}}$, is determined by minimizing the SCP free energy $\mathcal{F}_1$. 
We later show that the effect of this difference is negligible in Appendix \ref{Appendix_subsec_QHA_SCP_volume_diff}.

\subsubsection{The SCP frequency}
\label{subsec_SCP_freq}
From the SCP equation, the lowest order term of $\Omega_{\bm{k}\lambda} - \omega_{\bm{k}\lambda}$ is 
\begin{align}
\Omega_{\bm{k}\lambda} - \omega_{\bm{k}\lambda} 
\simeq
\frac{1}{N} \frac{1}{2\omega_{\bm{k}\lambda}} \sum_{\bm{k}'\lambda'} 
\widetilde{\Phi}(\bm{k}\lambda,-\bm{k}\lambda, \bm{k}'\lambda',-\bm{k}'\lambda')
g(\omega_{\bm{k}'\lambda'}),
\label{eq_Omega_omega_lowest_order1}
\end{align}
where $g(\omega) = \frac{\hbar}{2\omega} ( n_B(\hbar \omega) + \frac{1}{2} )$.
Because the lowest order terms of the expectation values of the harmonic and the quartic terms of the potential are
\begin{align}
\braket{\hat{U}_2} \simeq \sum_{\bm{k}\lambda} \frac{1}{2} \hbar \omega_{\bm{k}\lambda} \Bigl( n_B(\hbar \omega_{\bm{k}\lambda}) + \frac{1}{2} \Bigr)
\end{align}
\begin{align}
  \braket{\hat{U}_4} 
  \simeq& 
  \sum_{\bm{k}\lambda} \Bigl[ \frac{1}{2} \Bigl( n_B(\hbar \omega_{\bm{k}\lambda}) + \frac{1}{2} \Bigr) 
  \nonumber
  \\&
  \times
  \frac{1}{N}\frac{\hbar}{2 \omega_{\bm{k}\lambda}}
   \sum_{\bm{k}'\lambda'}  \widetilde{\Phi}(\bm{k}\lambda,-\bm{k}\lambda, \bm{k}'\lambda',-\bm{k}'\lambda')
   g(\omega_{\bm{k}'\lambda'})\Bigr],
  \nonumber
\end{align}
we get the average estimate of the frequency difference as 
\begin{align}
  \Omega_{\bm{k}\lambda} - \omega_{\bm{k}\lambda} \simeq \omega_{\bm{k}\lambda} \times \frac{\braket{\hat{U}_4}}{\braket{\hat{U}_2}}.
  \label{eq_Omega_minus_omega_lowest}
\end{align}
This is a reasonable estimate because the sign of $\Omega_{\bm{k}\lambda} - \omega_{\bm{k}\lambda}$ 
are the same for all the phonon modes in each material which we calculated.
The next dominant contribution to $\Omega_{\bm{k}\lambda} - \omega_{\bm{k}\lambda}$ can be calculated as
\begin{align}
  \Omega_{\bm{k}\lambda} - \omega_{\bm{k}\lambda} = \omega_{\bm{k}\lambda} \times \Bigl[ \frac{\braket{\hat{U}_4}}{\braket{\hat{U}_2}} + O\Bigl( \frac{\braket{\hat{U}_4}^2}{\braket{\hat{U}_2}^2}\Bigr) + \frac{3\braket{\hat{U}_6}}{2\braket{\hat{U}_2}} + \cdots \Bigr]
  \label{eq_Omega_minum_omega}
\end{align}
in the same manner, which is shown in Appendix \ref{Appendix_subsec_SCP_freq}.

\subsubsection{The Gr\"uneisen parameter, the mode specific heat, and the thermal expansion coefficient}
\label{subsec_perturbation_gamma_c_alpha}
Using the result of $\Omega_{\bm{k}\lambda} - \omega_{\bm{k}\lambda}$, we can perform the perturbation expansion for other quantities because the SCP free energy can be calculated from $\Omega_{\bm{k}\lambda}$.
Differentiating the SCP equation [Eq.~(\ref{eq_SCP_equation})] by the system volume $V$, we calculate the anharmonic correction to the Gr\"uneisen parameter $\gamma_{\bm{k}\lambda}$ as
\begin{equation}
  \gamma_{\bm{k}\lambda} \simeq \gamma^{\text{QHA}}_{\bm{k}\lambda} \times \Bigl[ 1 -(2+C) \frac{\braket{\hat{U}_4}}{\braket{\hat{U}_2}}  +  \frac{P_4}{P_2} \Bigr]
  \label{eq_gamma_correction}
\end{equation}
where $\gamma^{\text{QHA}}_{\bm{k}\lambda}$ is the Gr\"uneisen parameter calculated in QHA. 
The detailed derivation of this result is explained in Appendix \ref{Appendix_subsec_Gruneisen_param_alpha}.
$C$ is weakly temperature-dependent and takes the value of $1\sim2$. $P_2$ and $P_4$ are defined as
\begin{align}
  P_2 = \frac{1}{2} \times \sum_{\bm{k}\lambda} - \hbar \frac{\partial \omega_{\bm{k}\lambda}}{\partial V} \Bigl( n_B(\hbar \Omega_{\bm{k}\lambda}) + \frac{1}{2} \Bigr)
\end{align}
\begin{align}
  P_4 
  \simeq&
  \frac{-1}{2N} \sum_{\bm{k}\lambda,\bm{k}'\lambda'}  \frac{\partial \widetilde{\Phi}(\bm{k}\lambda,-\bm{k}\lambda,\bm{k}'\lambda',-\bm{k}'\lambda')}{\partial V}
  g(\Omega_{\bm{k}\lambda})
  g(\Omega_{\bm{k}'\lambda'}).
\end{align}
We call these terms as $P_2$ and $P_4$ because they are related to the SCP pressure, which we explain in Appendix \ref{Appendix_subsec_Gruneisen_param_alpha}.
We then calculate the correction to the $c_{v,\bm{k}\lambda}$. At high temperature, 
\begin{equation}
  c_{v,\bm{k}\lambda} 
  \simeq
  k_{\mathrm{B}} \Bigl[1 - \frac{1}{12}\Bigl( \frac{\hbar \Omega_{\bm{k}\lambda}}{k_{\mathrm{B}} T} \Bigr)^2 \Bigr].
\end{equation}
Thus
\begin{align}
  \Delta c_{v,\bm{k}\lambda} 
  &=
  c_{v,\bm{k}\lambda} -c^{\text{QHA}}_{v,\bm{k}\lambda} 
  \nonumber\\
  &\simeq
  - \frac{1}{6}k_{\mathrm{B}} \Bigl( \frac{\hbar \omega_{\bm{k}\lambda}}{k_{\mathrm{B}} T} \Bigr)^2 \frac{\Delta \omega_{\bm{k}\lambda}}{\omega_{\bm{k}\lambda}}
  \nonumber
  \\&\simeq
  - c^{\text{QHA}}_{v,\bm{k}\lambda} \times \frac{1}{6} \Bigl( \frac{\hbar \omega_{\bm{k}\lambda}}{k_{\mathrm{B}} T} \Bigr)^2 \frac{\braket{\hat{U}_4}}{\braket{\hat{U}_2}}.
\end{align}
In the low-temperature range of $T < \Theta_D$, where $\Theta_D$ is the Debye temperature, the modes with high frequency ($k_{\mathrm{B}} T \ll \hbar \omega_{\bm{k}\lambda}$) do not contribute to the thermal properties. Thus, we can consider that  
\begin{equation}
  \Delta c_{v,\bm{k}\lambda} 
  \simeq
  - c^{\text{QHA}}_{v,\bm{k}\lambda} \times \frac{1}{6} \frac{\braket{\hat{U}_4}}{\braket{\hat{U}_2}}
  \label{eq_cvklambda_correction}
\end{equation}
holds for the modes that contribute to the thermal properties.
The anharmonic correction to the bulk modulus can be neglected when considering the correction to the thermal expansion coefficient, which we discuss in Appendix \ref{Appendix_subsec_Gruneisen_param_alpha}.

Substituting Eqs.~(\ref{eq_gamma_correction}) and (\ref{eq_cvklambda_correction}) into the Gr\"uneisen formula [Eq.~(\ref{eq:alpha_gruneisen_SCP})], we get
\begin{equation}
    \alpha \simeq \alpha^{\text{QHA}}\times \Bigl( 1 - (\frac{13}{6}+C)\frac{\braket{\hat{U}_4}}{\braket{\hat{U}_2}} +\frac{P_4}{P_2} \Bigr).
    \label{eq_alpha_perturbation}
\end{equation}
In a fairly rough estimation, 
$P_4/P_2 \sim O({\braket{\hat{U}_4}}/{\braket{\hat{U}_2}})$
because $P_2$ and $P_4$ are proportional to the derivatives of the harmonic and the quartic IFCs in the definition. 
Therefore, QHA gives an accurate result for the thermal expansion coefficient if 
$| {\braket{\hat{U}_4}}/{\braket{\hat{U}_2}} | \ll1$, which weakly anharmonic materials satisfy at ambient conditions.

\begin{table*}[t]
  \caption{The summary of the analytical formulae and the results of the perturbation expansion of QHA and SCP that are derived in section \ref{sec_theory}.}
  \label{table_summary_formulae}
  \centering
  \begin{ruledtabular}
  {\renewcommand{\arraystretch}{2}
  \begin{tabular}{ccc}
       & QHA & SCP\\
      \hline
      Gr\"uneisen parameter 
      &
      $\gamma^{\text{QHA}}_{\bm{k}\lambda} = - \frac{V}{\omega_{\bm{k}\lambda}(V)} \Bigl( \frac{d \omega_{\bm{k}\lambda}(V)}{d V} \Bigr)$
      &
    $\gamma_{\bm{k}\lambda} = - \frac{V}{\Omega_{\bm{k}\lambda}(V,T)} \Bigl( \frac{d \Omega_{\bm{k}\lambda}(V,T)}{d V} \Bigr)
    \simeq  \gamma^{\text{QHA}}_{\bm{k}\lambda} \times \Bigl[ 1 -(2+C) \frac{\braket{\hat{U}_4}}{\braket{\hat{U}_2}}  +  \frac{P_4}{P_2} \Bigr]$
    \\ 
    Mode specific heat
    &
    $ c^{\text{QHA}}_{v,\bm{k}\lambda} = \hbar  \omega_{\bm{k}\lambda} \frac{\partial  n_B(\hbar \omega_{\bm{k}\lambda} )}{\partial T}$
    &
    $ c_{v,\bm{k}\lambda} = \hbar  \Omega_{\bm{k}\lambda}
    \Bigl(\frac{\partial  n_B(\hbar \Omega_{\bm{k}\lambda} )}{\partial T}\Bigr)_{\Omega} 
    \simeq 
    c^{\text{QHA}}_{v,\bm{k}\lambda} \times
    \Bigl( 1 - \frac{1}{6}\frac{\braket{\hat{U}_4}}{\braket{\hat{U}_2}} \Bigr)$
    \\ 
      Thermal expansion coefficient 
      & 
      $\alpha^{\text{QHA}} = \frac{1}{B^{\text{QHA}}_T} \frac{1}{V}\sum_{\bm{k}\lambda} c^{\text{QHA}}_{v,\bm{k}\lambda} \gamma^{\text{QHA}}_{\bm{k}\lambda} $ 
      & 
      $\alpha = \frac{1}{B_T} \frac{1}{V}\sum_{\bm{k}\lambda} c_{v,\bm{k}\lambda} \gamma_{\bm{k}\lambda} 
      \simeq\alpha^{\text{QHA}}\times \Bigl( 1 - (\frac{13}{6}+C)\frac{\braket{\hat{U}_4}}{\braket{\hat{U}_2}} +\frac{P_4}{P_2} \Bigr)$
      \\
      Phonon frequency shift 
      & 
      $\frac{\partial \omega_{\bm{k}\lambda}}{\partial T} = - \alpha^{\text{QHA}} \omega_{\bm{k}\lambda} \gamma^{\text{QHA}}_{\bm{k}\lambda}$
      &
    $\Bigl(\frac{\partial \Omega_{\bm{k}\lambda}^{\text{Hes}}}{\partial T} \Bigr)_{P} = - \alpha \Omega_{\bm{k}\lambda} \gamma_{\bm{k}\lambda} + \left( \frac{\partial \Omega_{\bm{k}\lambda}}{\partial T} \right)_{V} + \Bigl( \frac{\partial \Delta \Omega^{\text{Hes}}_{\bm{k}\lambda}}{\partial T} \Bigr)_{P}$
    \\ 
  \end{tabular}
  }
  \end{ruledtabular}
\end{table*}

\subsubsection{The temperature-dependent phonon frequency shift}
\label{subsubsec_phonon_freq_shift}
We have shown that the anharmonic corrections to $\alpha$ and $\gamma_{\bm{k}\lambda}$ are of order
$O({\braket{\hat{U}_4}}/{\braket{\hat{U}_2}})$. Thus, from Eq.~(\ref{eq_anharmonic_freq_shift}), the condition for QHA to give an accurate result of the phonon frequency shift is
\begin{equation}
    |\alpha \Omega_{\bm{k}\lambda} \gamma_{\bm{k}\lambda} |\gg 
   \Bigl| \Bigl( \frac{\partial \Omega_{\bm{k}\lambda}}{\partial T} \Bigr)_{V} 
    +
    \Bigl( \frac{\partial \Delta \Omega^{\text{Hes}}_{\bm{k}\lambda}}{\partial T} \Bigr)_P \Bigr|
    \label{eq_QHA_freq_condition}
\end{equation}
There is no guarantee that the two terms in the RHS of Eq.~(\ref{eq_QHA_freq_condition}) cancel out each other because
$({\partial \Delta \Omega^{\text{Hes}}_{\bm{k}\lambda}}/{\partial T} )_P$ is always negative 
while the sign of 
$( {\partial \Omega_{\bm{k}\lambda}}/{\partial T} )_{V} $
depends on materials.
Therefore, for QHA to accurately reproduce the $T$-dependent phonon frequency shift not by an accidental cancellation,
the system must satisfy both
\begin{align}
&|\alpha \Omega_{\bm{k}\lambda} \gamma_{\bm{k}\lambda} |\gg 
   \Bigl| \Bigl( \frac{\partial \Omega_{\bm{k}\lambda}}{\partial T} \Bigr)_{V} \Bigr|
   \label{eq_QHA_freq_condition1}
   \\
&|\alpha \Omega_{\bm{k}\lambda} \gamma_{\bm{k}\lambda} |\gg 
    \Bigl|\Bigl( \frac{\partial \Delta \Omega^{\text{Hes}}_{\bm{k}\lambda}}{\partial T} \Bigr)_P \Bigr|
    \label{eq_QHA_freq_condition2}
\end{align}
The order estimation of the phonon frequency shift is difficult because it is not a thermodynamic quantity, so we focus on the condition of Eq.~(\ref{eq_QHA_freq_condition1}) and show that it is hard to satisfy for ordinary harmonic materials.
As for the condition of Eq.~(\ref{eq_QHA_freq_condition2}), we note that the magnitude of 
$| ( {\partial \Omega_{\bm{k}\lambda}}/{\partial T} )_{V} |$
and 
$|( {\partial \Delta \Omega^{\text{Hes}}_{\bm{k}\lambda}}/{\partial T} )_P |$
are comparable according to the results of our first-principles calculation. For silicon and diamond, the two terms of the RHS of Eq.~(\ref{eq_QHA_freq_condition}) have the same sign and $|\alpha \Omega_{\bm{k}\lambda} \gamma_{\bm{k}\lambda} |$ was smaller than
$
   | ( {\partial \Omega_{\bm{k}\lambda}}/{\partial T} )_{V} 
    +
    ( {\partial \Delta \Omega^{\text{Hes}}_{\bm{k}\lambda}}/{\partial T} )_P |
$. As a result, QHA gives a bad estimate of the $T$-dependent phonon frequency shift for these materials.
For NaCl and MgO, $ ( {\partial \Omega_{\bm{k}\lambda}}/{\partial T} )_{V} $  was positive,
for which the positivity of $\braket{\hat{U}_4}$ is important because of Eq.~(\ref{eq_Omega_minus_omega_lowest}) and because $\Omega_{\bm{k}\lambda}(V,T) - \omega_{\bm{k}\lambda}(V)$ is roughly proportional to the temperature. For these materials, an accidental cancellation occur in the RHS of Eq.~(\ref{eq_QHA_freq_condition}) and QHA accidentally gives a good result for the phonon frequency shift even though the conditions Eqs.~(\ref{eq_QHA_freq_condition1}) and (\ref{eq_QHA_freq_condition2}) are not satisfied.
Thus, QHA seems to produce a better result for the phonon frequency shift than for the thermal expansion coefficient because $| {\braket{\hat{U}_4}}/{\braket{\hat{U}_2}} |$ of NaCl and MgO are not so small.
We infer this applies to many materials with strong anharmonicity because $\braket{\hat{U}_4}$ is usually positive in these materials, which makes them stable when the bottom of the potential well is nearly flat.
For the order estimation of Eq.~(\ref{eq_QHA_freq_condition1}),
we use the following typical orders of physical quantities 
\begin{align}
  &\gamma_{\bm{k}\lambda} \sim O(1),
  \label{eq_order_gamma} \\ 
  &\alpha \sim 10^{-5}\text{--}10^{-6}\text{\ K}^{-1}.
  \label{eq_order_alpha}
\end{align}
The lowest order contribution to 
$| ( {\partial \Omega_{\bm{k}\lambda}}/{\partial T} )_{V} |$ is estimated as
\begin{align}
&
  \Bigl( \frac{ \partial \Omega_{\bm{k}\lambda} }{\partial T}\Bigr)_{V} 
  \nonumber
  \\
  &\simeq
  \frac{1}{N} \sum_{\bm{k}'\lambda'} \frac{\hbar}{4} \frac{\widetilde{\Phi}(\bm{k}\lambda,-\bm{k}\lambda,\bm{k}'\lambda',-\bm{k}'\lambda')}{\omega_{\bm{k}\lambda}\omega_{\bm{k}'\lambda'}} \frac{\partial}{\partial T}\Bigl( n_B(\hbar \Omega_{\bm{k}_1\lambda_1}) + \frac{1}{2} \Bigr)
  \nonumber
  \\&\sim
  \frac{1}{T}\frac{1}{N} \sum_{\bm{k}'\lambda'} \frac{\hbar}{4} \frac{\widetilde{\Phi}(\bm{k}\lambda,-\bm{k}\lambda,\bm{k}'\lambda',-\bm{k}'\lambda')}{\omega_{\bm{k}\lambda}\omega_{\bm{k}'\lambda'}} \Bigl( n_B(\hbar \Omega_{\bm{k}_1\lambda_1}) + \frac{1}{2} \Bigr)
  \nonumber
  \\&\sim
  \frac{\omega_{\bm{k}\lambda}}{T} \times \frac{\braket{\hat{U}_4}}{\braket{\hat{U}_2}},
  \label{eq_partial_Omega_partial_T}
\end{align}
Substituting Eqs.~(\ref{eq_order_gamma})--(\ref{eq_partial_Omega_partial_T})
into Eq.~(\ref{eq_QHA_freq_condition1}), 
the condition that Eq.~(\ref{eq_QHA_freq_condition1}) is satisfied is
\begin{align}
  \Bigl| \frac{\braket{\hat{U}_4}}{\braket{\hat{U}_2}} \Bigr| \ll 10^{-3} \sim 10^{-4}
\end{align}
at $T\simeq 10^2$ K. This condition is much more strict than the condition for QHA to be an accurate approximation for the thermal expansion and is not satisfied even by weakly anharmonic materials such as silicon.

The discussion on the accuracy of QHA is summarized in Table \ref{table_summary_QHA_applicable_limit}, in conjunction with the analysis of the numerical calculation in the later section.

\begin{table*}[t]
  \caption{The range of applicability of the QHA.}
  \label{table_summary_QHA_applicable_limit}
  \centering
  \begin{ruledtabular}
  
  \begin{tabular}{ccc}
      & Weakly harmonic materials & Strongly anharmonic materials\\
      \hline
      Thermal expansion coefficient 
      &
      always good
      &
      deviation of $O(\frac{\braket{\hat{U}_4}}{\braket{\hat{U}_2}})$
    \\ 
    Phonon frequency shift
    &
    poor when $\braket{\hat{U}_4}<0$
    & 
    empirically good because $\braket{\hat{U}_4}$ is usually positive
    \\ 
  \end{tabular}
    \end{ruledtabular}
\end{table*}
\section{Simulation Methods}
\subsection{QHA and SCP calculation of thermal expansion}
We perform the first-principles calculation on silicon, diamond (covalent crystals), NaCl (ionic crystal) and MgO (oxide). We use the \textit{Vienna ab inito simulation package} (VASP)~\cite{PhysRevB.54.11169} for the electronic structure calculation and ALAMODE~\cite{Tadano_2014, PhysRevB.92.054301, PhysRevMaterials.3.033601} for calculating the phonon properties.

In the QHA calculation, we calculate the harmonic phonon dispersion by using the small displacement method.
We displace the atoms from their equilibrium positions by 0.01 \AA\ and extract the harmonic IFCs from the force-displacement patterns by using the least-square method implemented in the ALAMODE package~\cite{Tadano_2014}. 
The free energy is calculated by using Eq.~(\ref{eq_QHA_free_energy}) for 10--20 different lattice constants, and the optimum lattice constant is determined by fitting the free energy by the Birch--Murnaghan equation of state~\cite{PhysRev.71.809, Murnaghan244} at each temperature.

For the SCP calculation, we truncate the Taylor expansion of the PES at the quartic order and fit the potential by $\hat{U}_2 + \hat{U}_3 + \hat{U}_4$. 
We fix the harmonic IFCs at the values calculated by the small displacement method and optimize the cubic and quartic IFCs. 
The optimization is performed by the compressive sensing method~\cite{PhysRevLett.113.185501, PhysRevB.92.054301}, which enables us to efficiently extract the IFCs from a small number of displacement patterns. 
We run the SCP calculation and calculate the SCP free energy for several lattice constants, which is again fitted by the Birch-Murnaghan equation of state to determine the optimum lattice constant at each temperature.

The VASP package is used for the force calculations of the displacement patterns and for calculating the electronic ground state energies.

\subsection{Expectation values of the harmonic and quartic terms in the potential energy surface}
We calculate the $T$-dependence of 
${\braket{\hat{U}_4}}/{\braket{\hat{U}_2}}$.
As the lowest-order approximation, the expectation value is taken with respect to the density matrix of the harmonic potential $\hat{U}_2$.
$\braket{\hat{U}_2}$ is calculated using the analytic formula from the harmonic phonon dispersion. ${\braket{\hat{U}_4}}$ is estimated using a stochastic method. We generate random configurations that obey the density matrix of the harmonic Hamiltonian~\cite{PhysRevB.89.064302}, and calculate the DFT energy for each configuration. Assuming that the contribution from the higher-order terms are negligible, ${\braket{\hat{U}_4}}$ is estimated as
\begin{equation}
{\braket{\hat{U}_4}} \simeq \frac{1}{N_i}\sum_{i\text{: configurations}} (U_i - U_{\text{harm},i})
\end{equation}
where $U_i$ and $U_{\text{harm},i}$ are the DFT energy and the energy in the harmonic approximation of the potential of the $i$-th configuration respectively.
To remove the contributions
from the odd-order terms of the PES, the configurations are generated so that the displacements in the $2i$-th configuration and those in the $2i+1$-th configuration are in the opposite direction with the same magnitude.

\subsection{Simulation details}
We use $2\times 2\times 2$ cubic supercell, which is generated from the conventional cell, for the phonon calculation. The supercell of each material contains 64 atoms. 
In order to check the convergence with respect to the supercell size,
we also calculated the harmonic IFCs, which tend to be more long-ranged than the anharmonic IFCs, using a $3\times 3\times 3$ supercell. The calculation results of the harmonic phonon dispersion are well converged for all the materials we calculate.

For the electronic structure calculations, we use the VASP implementation of the PBEsol exchange-correlation functional~\cite{PhysRevLett.100.136406} and PAW pseudopotentials~\cite{PhysRevB.50.17953,PhysRevB.59.1758}.
For NaCl and MgO, the Born effective charge is calculated by  density functional perturbation theory~\cite{PhysRevB.33.7017, PhysRevB.73.045112} to get the nonanalytic part of the dynamical matrix.
The Brillouin zone integration is performed over the $4\times4\times 4$ Monkhorst-Pack $k$-mesh, which we checked is sufficient for the convergence of the total energy.
We set the convergence criteria of the SCF loop as $10^{-8}$ eV and use the accurate precision mode, which reduces egg-box effects and errors. The basis cutoff we use are 500 eV for silicon and 600 eV for diamond, MgO, and NaCl.

\begin{figure}[!b]
\centering
\includegraphics[width=0.48\textwidth]{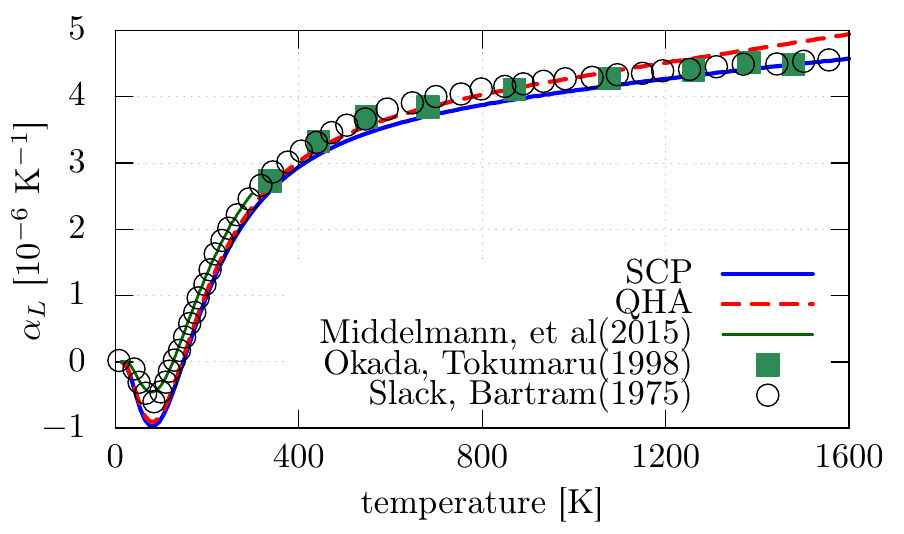}
\caption{
Linear thermal expansion coefficient $\alpha_L$ of silicon calculated by SCP and QHA.
The experimental results are taken from Refs.~\cite{PhysRevB.92.174113, doi:10.1063/1.333965, doi:10.1063/1.321373}
}
\label{fig_Si_T_alpha_L}
\end{figure}

\section{Result and Discussion}
\label{sec_results}
\subsection{silicon} 
\label{subsec_silicon}

\begin{figure}[tb]
\centering
\includegraphics[width=0.48\textwidth]{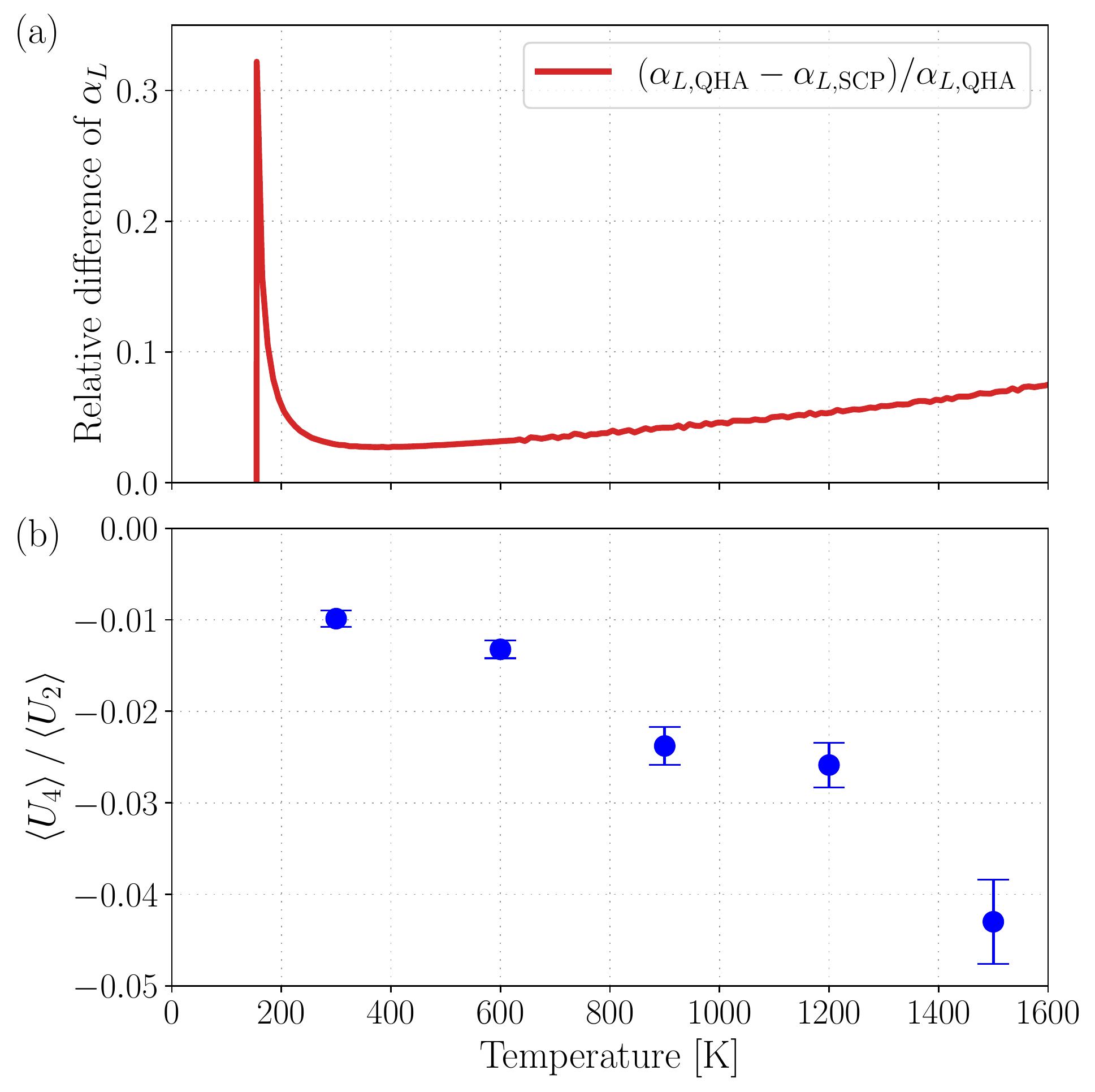}

\caption{
(a) Relative difference of the QHA result and the SCP result of the linear thermal expansion coefficient $\alpha_L$ of silicon. (b) The calculation result of 
${\braket{\hat{U}_4}}/{\braket{\hat{U}_2}}$ of silicon. 
}\label{fig_Si_reldiff_U4U2}
\end{figure}

The calculation result of the thermal expansion coefficient of silicon is shown in Fig.~\ref{fig_Si_T_alpha_L}. We find that both the SCP result and the QHA result are in reasonable agreement with the experimental results. It should be noted that we plot the linear thermal expansion coefficient 
$\alpha_L = \frac{1}{a} ({\partial a}/{\partial T})_P$ for direct comparison with the experiments. The linear thermal expansion coefficient can be written as 
$\alpha_L = \alpha/3$ for isotropic materials, where $\alpha$ is the volume thermal expansion coefficient.
In Fig.~\ref{fig_Si_reldiff_U4U2}, we compare 
$({\alpha_{L,\text{SCP}} - \alpha_{L,\text{QHA}}})/{\alpha_{L,\text{QHA}}}$ and
${\braket{\hat{U}_4}}/{\braket{\hat{U}_2}}$ of silicon.
The spiky structure in Fig.~\ref{fig_Si_reldiff_U4U2}(a) occurs because the thermal expansion coefficient changes its sign at $\sim$200 K. At higher temperatures,
$({\alpha_{L,\text{SCP}} - \alpha_{L,\text{QHA}}})/{\alpha_{L,\text{QHA}}}$ and
${\braket{\hat{U}_4}}/{\braket{\hat{U}_2}}$ are in the same order, which is consistent with our theoretical considerations [Eq.~(\ref{eq_alpha_perturbation})].

\begin{figure}[tb]
\centering
\includegraphics[width=0.48\textwidth]{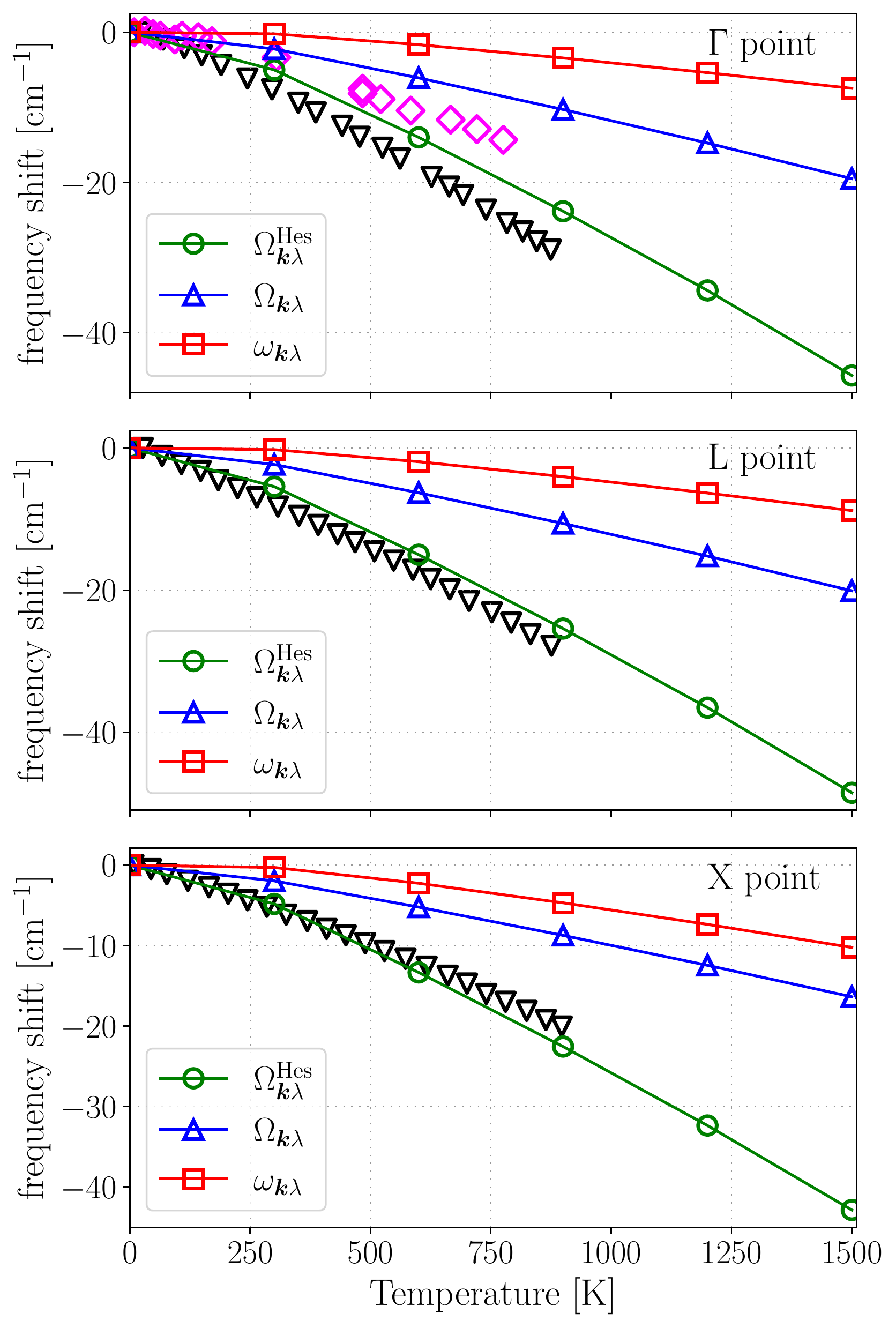}

\caption[]{
Temperature-dependent frequency shift of the TO modes of silicon at
the $\Gamma$ point (top), the L point (middle), and the X point (bottom).
The frequency shift is measured from the frequency at zero temperature in the same calculation method or experiment. 
$\Omega^\text{Hes}_{\bm{k}\lambda}$ and $\Omega_{\bm{k}\lambda}$ are defined by Eq.~(\ref{eq_def_Hessian_freq}) and Eq.~(\ref{eq_SCP_equation}), respectively. Note that $\Omega^\text{Hes}_{\bm{k}\lambda}$ is considered to correspond to the experimentally measured phonon frequency.
$\omega_{\bm{k}\lambda}$ is the QHA frequency, which is explained in section \ref{subsec_the_Gruneisen_theory}.
The experimental data is taken from Ref.~\cite{doi:10.1063/1.93394} (black open triangle) and  Ref.~\cite{PhysRevB.29.2051} (magenda open diamond).
}\label{fig_Si_optical_freq_shift}

\end{figure}

\begin{figure}[H]
\centering
\includegraphics[width=\columnwidth]{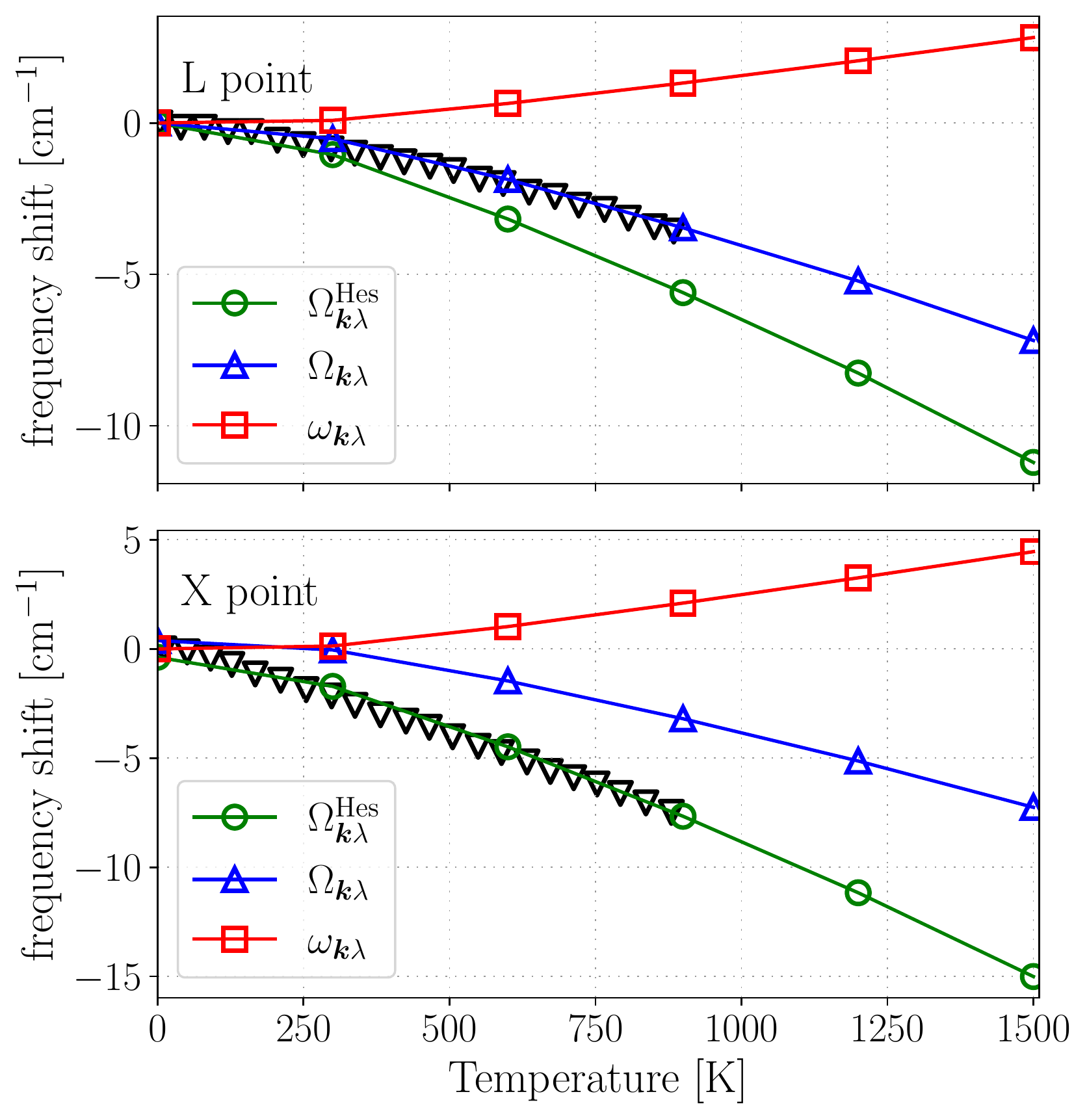}

\caption[]{
Temperature-dependent frequency shift of the TA modes of silicon 
at the L point (top) and the X point (bottom).
The frequency shift is measured from the frequency at zero temperature in the same method or experiment. 
$\Omega^\text{Hes}_{\bm{k}\lambda}$ and $\Omega_{\bm{k}\lambda}$are defined in Eq. (\ref{eq_def_Hessian_freq}), Eq. (\ref{eq_SCP_equation}) respectively. Note that $\Omega^\text{Hes}_{\bm{k}\lambda}$ is considered to correspond to the experimentally measured phonon frequency.
$\omega_{\bm{k}\lambda}$ is the QHA frequency, which is explained in section \ref{subsec_the_Gruneisen_theory}.
The experimental data is taken from Ref.~\cite{doi:10.1063/1.93394}.
}\label{fig_Si_TA_freq_shift}

\end{figure}

Figure \ref{fig_Si_optical_freq_shift} shows the frequency shift of the transverse optical (TO) phonon of silicon at high-symmetry points.
As the temperature rises, the SCP frequency of silicon softens more rapidly than the QHA frequency, which is consistent with the negativity of $\braket{\hat{U_4}}$ (see Fig.~\ref{fig_Si_reldiff_U4U2}(b)). $\Delta \Omega^{\text{Hes}}_{\bm{k}\lambda}$ further softens the frequency because $\Delta \Omega^{\text{Hes}}_{\bm{k}\lambda}$ is always negative and its magnitude gets larger at higher temperature. 
Consequently, the TO mode of silicon softens much faster than the QHA prediction, which was also pointed out in the previous research~\cite{PhysRevB.91.014307}.
We plot the frequency shift of the TA mode of silicon in Fig.~\ref{fig_Si_TA_freq_shift}. The QHA frequency gets larger when the temperature rises because its Gr\"uneisen parameter is negative. However, the experimentally measured frequency gets softer~\cite{Kim1992}. This discrepancy is resolved in
$\Omega_{\bm{k}\lambda}$ or $\Omega^{\text{Hes}}_{\bm{k}\lambda}$ which correctly consider the effect of phonon-phonon interaction.

\subsection{diamond}
\begin{figure}[h]
\vspace{0cm}
\begin{center}
\includegraphics[width=0.48\textwidth]{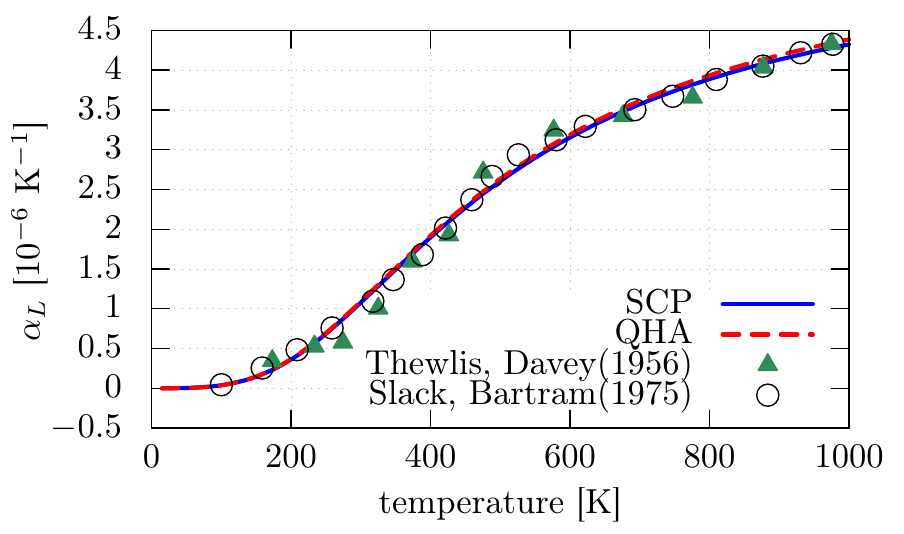}
\caption{
The linear thermal expansion coefficient $\alpha_L$ of diamond calculated by SCP and QHA.
The experimental results are taken from Refs.~\cite{doi:10.1080/14786435608238119, doi:10.1063/1.321373}
}
\label{fig_C_T_alpha_L}
\end{center}
\end{figure}

As shown in Fig.~\ref{fig_C_T_alpha_L}, SCP and QHA produce almost the same result for the thermal expansion coefficient, which agrees very well with the experimental results. 
This is because that the effect of the lattice anharmonicity is small in diamond. 

\begin{figure}[H]
\vspace{0cm}
\begin{center}
\includegraphics[width=0.48\textwidth]{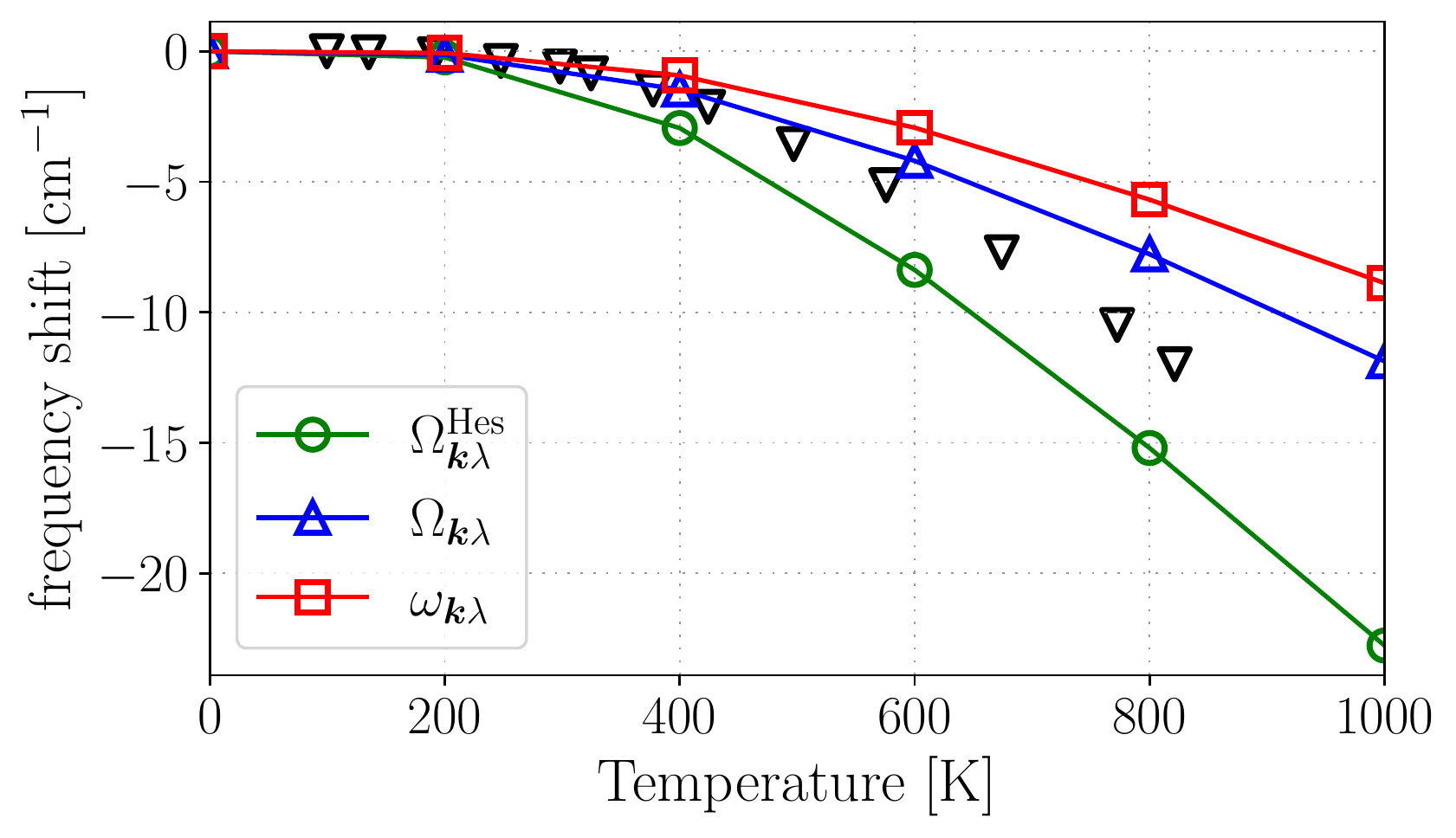}
\caption{
Temperature-dependent frequency shift of the optical mode of diamond at the $\Gamma$ point. The frequency shift is measured from the frequency at zero temperature in the same calculation method or experiment.
$\Omega^\text{Hes}_{\bm{k}\lambda}$ and $\Omega_{\bm{k}\lambda}$are defined in Eq.~(\ref{eq_def_Hessian_freq}) and Eq.~(\ref{eq_SCP_equation}), respectively. Note that $\Omega^\text{Hes}_{\bm{k}\lambda}$ is considered to correspond to the experimentally measured phonon frequency.
$\omega_{\bm{k}\lambda}$ is the QHA frequency, which is explained in section \ref{subsec_the_Gruneisen_theory}.
The experimental data is taken from Ref.~\cite{PhysRevB.61.3391}.
}
\label{fig_C_Gamma_TLO}
\end{center}
\end{figure}

Figure \ref{fig_C_Gamma_TLO} represents the $T$-dependent phonon frequency shift of diamond.
In diamond, both the $\Omega_{\bm{k}\lambda}-\omega_{\bm{k}\lambda}$ and $\Delta \Omega^{\text{Hes}}_{\bm{k}\lambda}$ are negative, which resembles the tendency in silicon. 
Although the agreement with the experiment is not perfect, the 
$\Omega_{\bm{k}\lambda}$ and $\Omega^{\text{Hes}}_{\bm{k}\lambda}$ explains the fact that the frequency softens more rapidly than the QHA result.

We also calculated the $T$-dependence of
${\braket{\hat{U}_4}}/{\braket{\hat{U}_2}}$ for diamond, but we could not obtain a meaningful result for $\braket{\hat{U}_4}$ within our method because $|\braket{\hat{U}_4}|$ was smaller than the statistical error.

\clearpage
\subsection{NaCl} 

\begin{figure}[h]
\vspace{0cm}
\begin{center}
\includegraphics[width=0.48\textwidth]{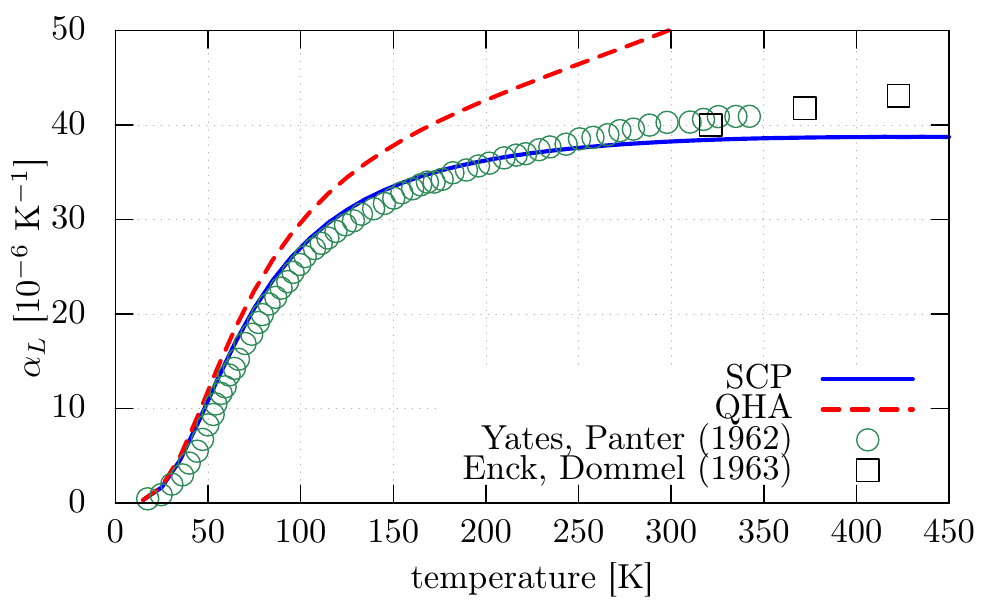}
\caption{
Linear thermal expansion coefficient $\alpha_L$ of NaCl calculated by SCP and QHA.
The experimental results are taken from Refs.~\cite{SRIVASTAVA19732069, Yates1962ThermalEO}
}
\label{fig_NaCl_T_alpha_L}
\end{center}
\end{figure}

\begin{figure}[H]
\centering
\includegraphics[width=0.48\textwidth]{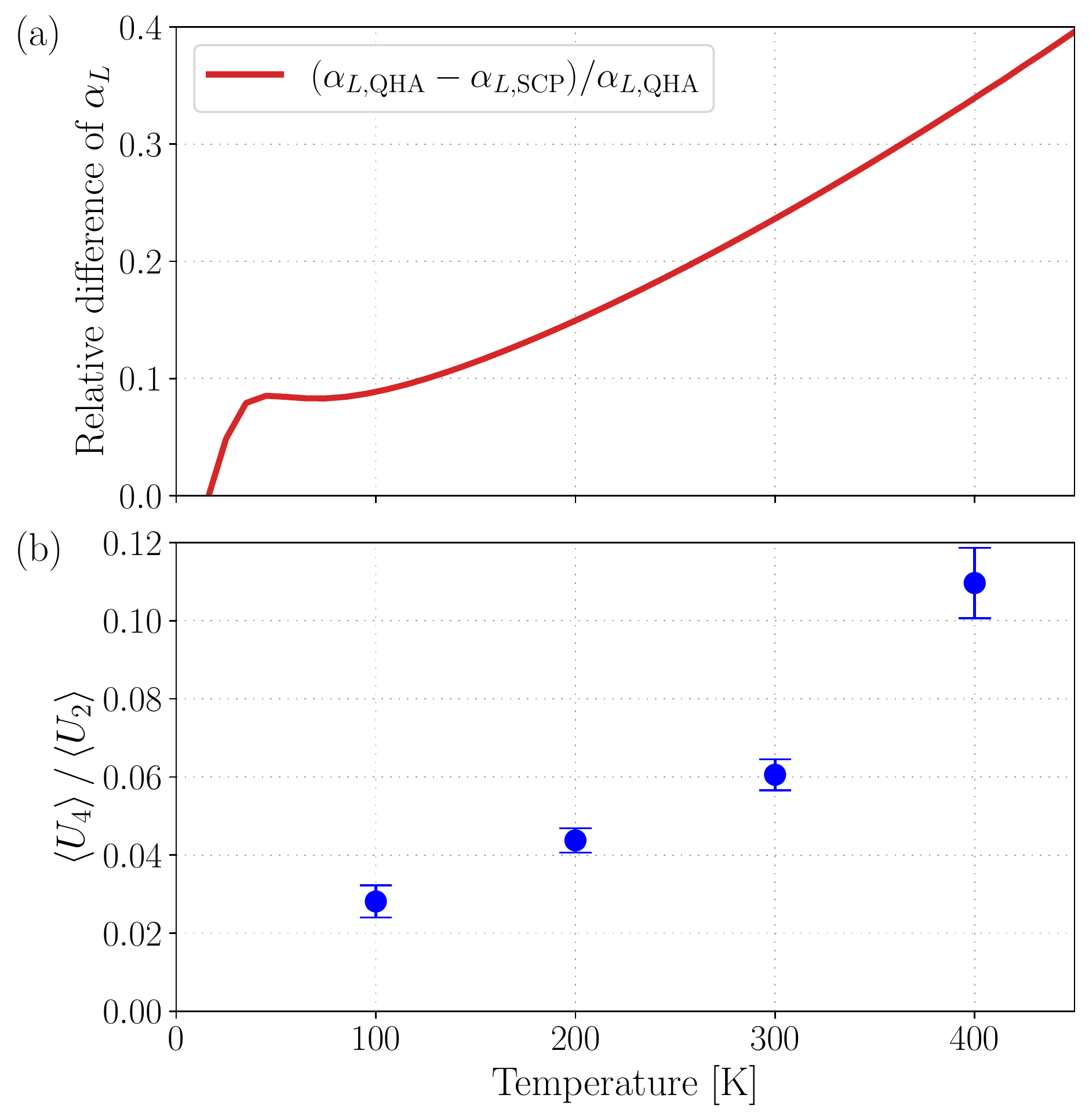}

\caption{
(a) Relative difference of the QHA result and the SCP result of the linear thermal expansion coefficient $\alpha_L$ of NaCl. (b) The calculation result of
${\braket{\hat{U}_4}}/{\braket{\hat{U}_2}}$ of NaCl.
}\label{fig_NaCl_reldiff_U4U2}

\end{figure}

As is depicted in Fig.~\ref{fig_NaCl_T_alpha_L}, there is a clear deviation between the thermal expansion coefficient calculated by QHA and SCP, which signifies that the lattice anharmonicity plays a significant role in NaCl. QHA overestimates the thermal expansion coefficient from low temperatures, but this trend is suppressed in SCP.
In a previous work by Ravichandran and Broido~\cite{PhysRevB.98.085205}, they additionally consider another term of the free energy, which consists of squared cubic IFCs. We consider that the volume dependence, which is essential for the term to affect the thermal expansion, of this term is small because our calculation result is consistent with theirs in the temperature range of our calculation.
In Fig.~\ref{fig_NaCl_reldiff_U4U2}, we can see that
$(\alpha_{L,\text{QHA}}-\alpha_{L,\text{SCP}})/\alpha_{L,\text{QHA}} \sim 3.3\times {\braket{\hat{U}_4}}/{\braket{\hat{U}_2}}$, which is in very good agreement with Eq.~(\ref{eq_alpha_perturbation}) when we assume that ${P_4}/{P_2}$ is relatively small.

\begin{figure}[H]
\vspace{0cm}
\begin{center}
\includegraphics[width=0.48\textwidth]{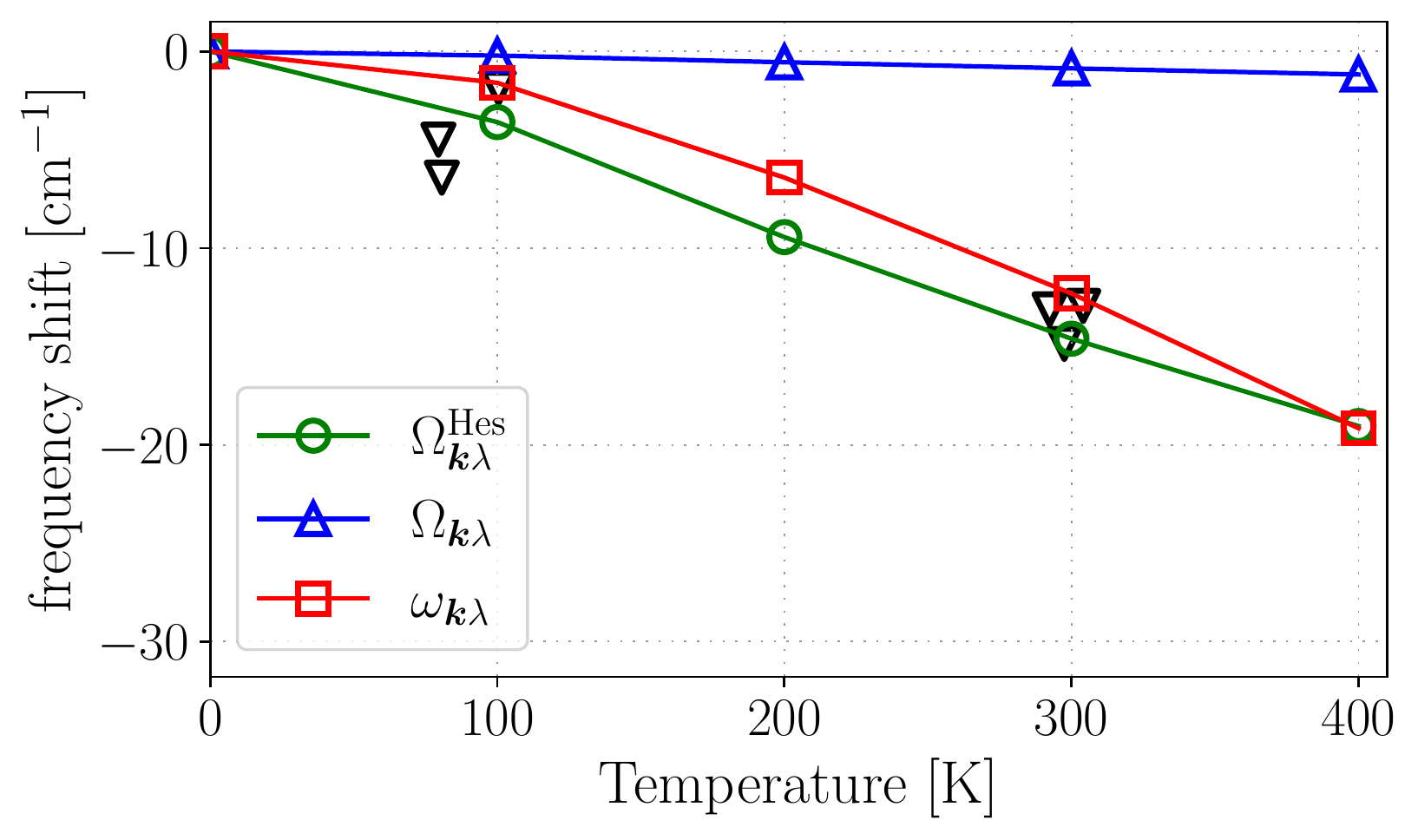}
\caption{
Temperature-dependent frequency shift of the transverse optical (TO) modes of NaCl at the $\Gamma$ point. The frequency shift is measured from the frequency at zero temperature in the same calculation method or experiment.
$\Omega^\text{Hes}_{\bm{k}\lambda}$ and $\Omega_{\bm{k}\lambda}$are defined in Eq.~(\ref{eq_def_Hessian_freq}) and Eq.~(\ref{eq_SCP_equation}), respectively. Note that $\Omega^\text{Hes}_{\bm{k}\lambda}$ is considered to correspond to the experimentally measured phonon frequency.
$\omega_{\bm{k}\lambda}$ is the QHA frequency, which is explained in section \ref{subsec_the_Gruneisen_theory}.
The experimental data is taken from Ref.
 \cite{Cowley_1972}.
}
\label{fig_NaCl_Gamma_TO}
\end{center}
\end{figure}

As for the phonon frequency shift, we plot the $T$-dependent frequency shift of the TO mode of NaCl at $\Gamma$ point in Fig.~\ref{fig_NaCl_Gamma_TO}.
We can see that 
$\Omega_{\bm{k}\lambda} - \omega_{\bm{k}\lambda}$ and $\Delta \Omega^{\text{Hes}}_{\bm{k}\lambda}$
cancel almost entirely with each other. As a result, the QHA result agrees well with the 
$\Omega^{\text{Hes}}_{\bm{k}\lambda}$ and accidentally explains the experimental trend.
This cancellation occurs because $\braket{\hat{U}_4}$ is positive, which makes $\Omega_{\bm{k}\lambda} - \omega_{\bm{k}\lambda}$ positive. For the detailed discussion on the sign of the phonon frequency shift, see section \ref{subsubsec_phonon_freq_shift}.
Consequently, the QHA works better for the phonon frequency shift than for the thermal expansion coefficient in this material. We consider that this trend is common to a wide range materials with strong anharmonicity because $\braket{\hat{U}_4}$ tend to be positive when the curvature of the PES at the potential minimum is small, which we present in Table \ref{table_summary_QHA_applicable_limit}.
However, special care must be taken when applying QHA to anharmonic materials because this cancellation is not theoretically ensured. In addition, this result shows that we cannot necessarily justify the use of QHA when the phonon frequency shift agrees with experiments.

\clearpage
\subsection{MgO}
\begin{figure}[htbp]
\vspace{0cm}
\begin{center}
\includegraphics[width=0.48\textwidth]{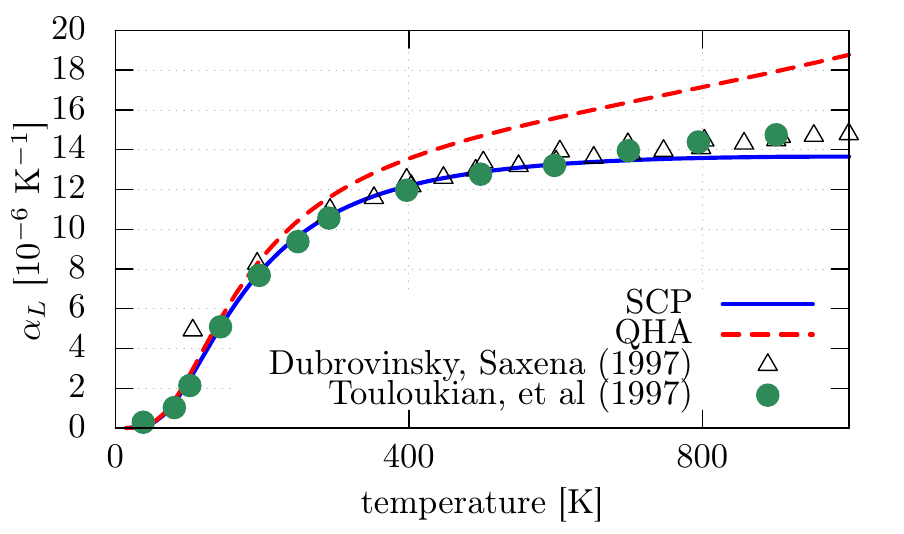}
\caption{
Calculation results of linear thermal expansion coefficient $\alpha_L$ of MgO compared with experimental results taken from Refs.~\cite{Dubrovinsky1997ThermalEO, touloukian1977thermal}
}
\label{fig_MgO_T_alpha_L}
\end{center}
\end{figure}

\begin{figure}[htbp]
\centering
\includegraphics[width=0.48\textwidth]{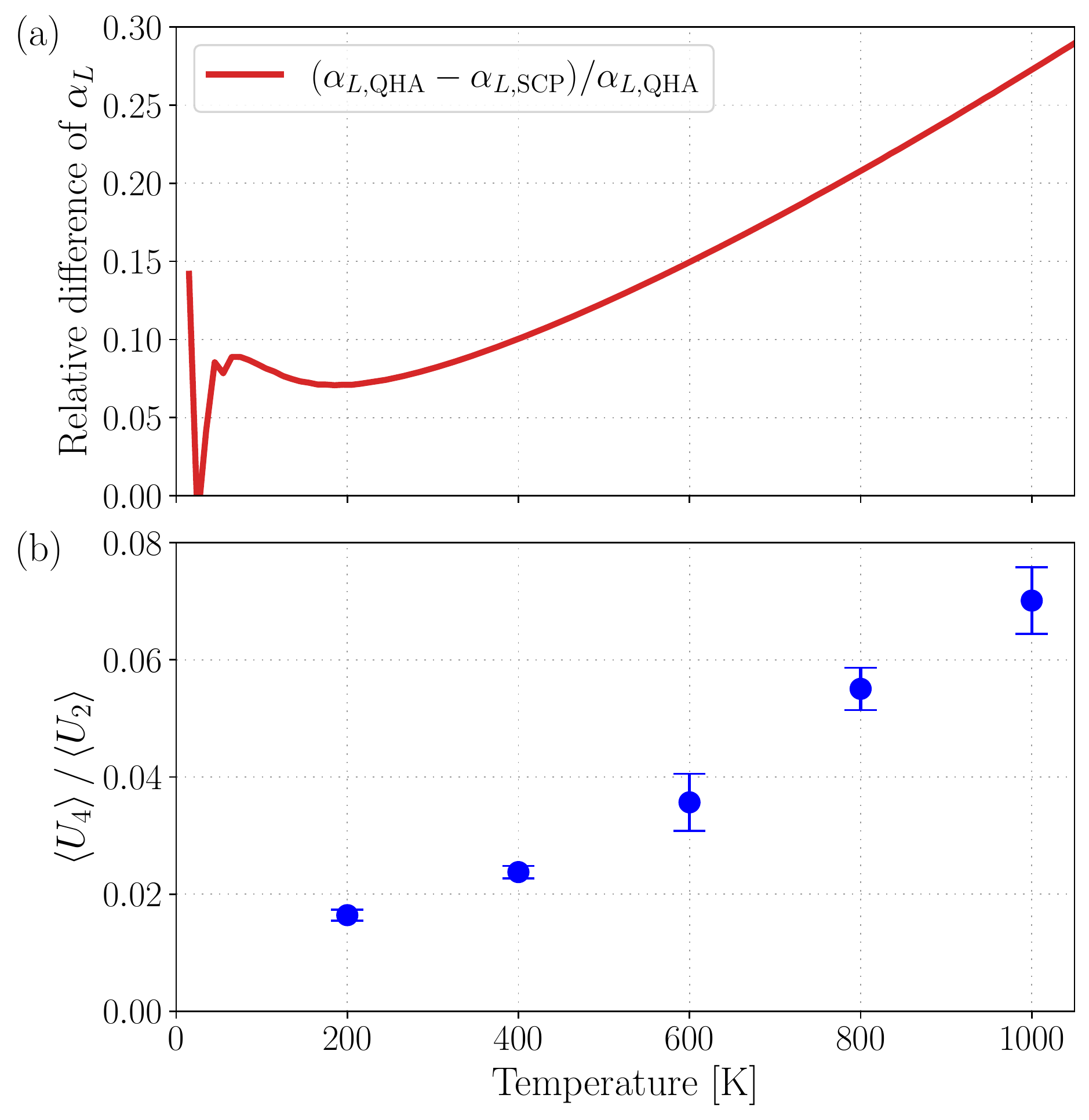}

\caption{
(a) The relative difference of the QHA result and the SCP result of the linear thermal expansion coefficient $\alpha_L$ of MgO. (b) The calculation result of
${\braket{\hat{U}_4}}/{\braket{\hat{U}_2}}$ of MgO.
}\label{fig_MgO_reldiff_U4U2}

\end{figure}

\begin{figure}[htbp]
\centering
\includegraphics[width=\columnwidth]{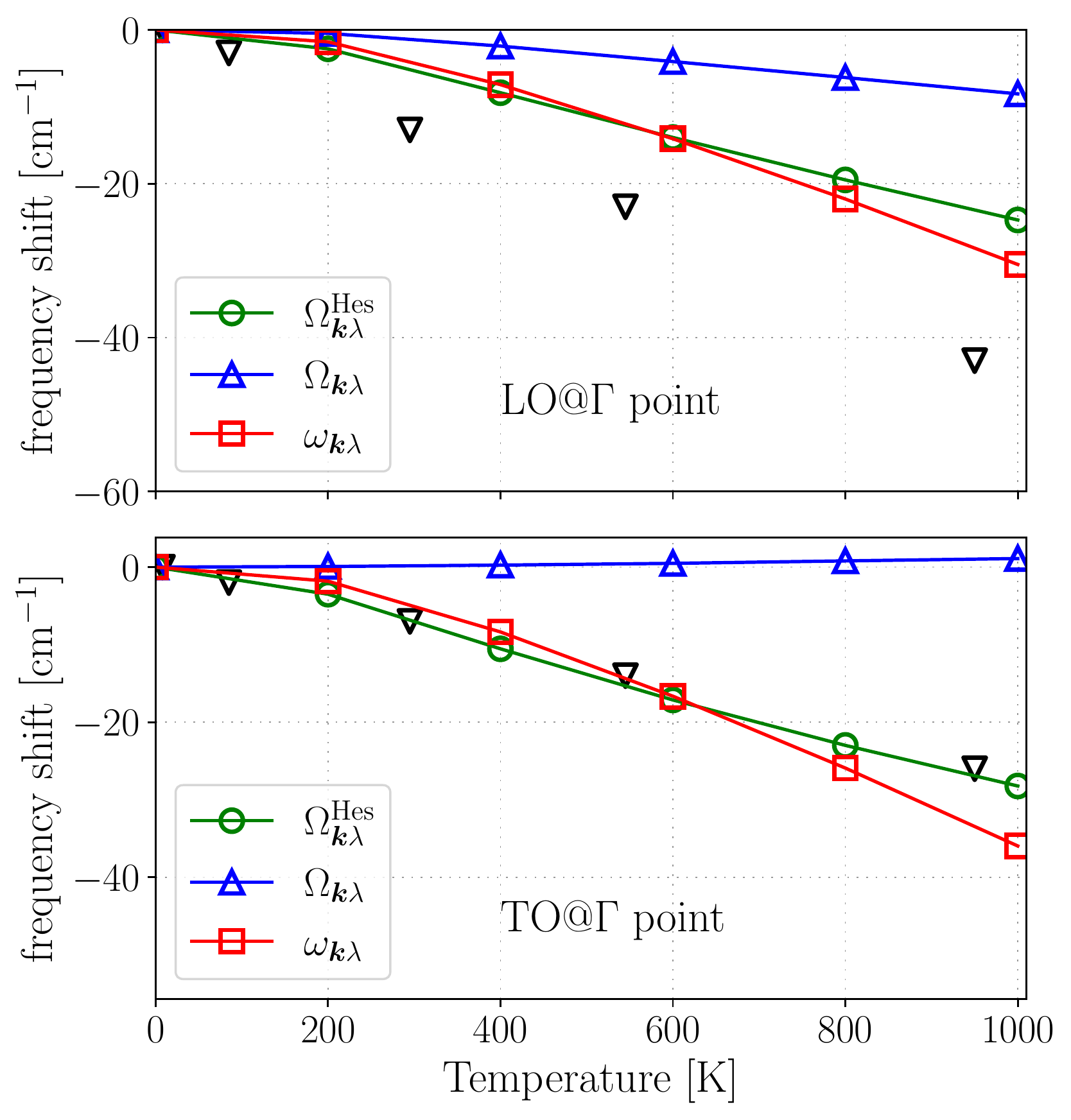}

\caption[]{
The temperature-dependent frequency shift of
the longitudinal optical (LO) mode (top panel) and the the transverse optical (TO) mode (bottom panel) 
of MgO at the $\Gamma$ point. The frequency shift is measured from the frequency at zero temperature in the same method or experiment.
$\Omega^\text{Hes}_{\bm{k}\lambda}$ and $\Omega_{\bm{k}\lambda}$are defined in Eq.~(\ref{eq_def_Hessian_freq}) and Eq.~(\ref{eq_SCP_equation}), respectively. Note that $\Omega^\text{Hes}_{\bm{k}\lambda}$ is considered to correspond to the experimentally measured phonon frequency.
$\omega_{\bm{k}\lambda}$ is the QHA frequency, which is explained in section \ref{subsec_the_Gruneisen_theory}.
The experimental data is taken from Ref.~\cite{PhysRev.146.526}.
}\label{fig_MgO_Gamma_optical_freq_shift}

\end{figure}

According to Fig.~\ref{fig_MgO_T_alpha_L}, the thermal expansion coefficient calculated by SCP agrees well with the experimental results while QHA overestimates the thermal expansion coefficient. From Fig.~\ref{fig_MgO_reldiff_U4U2}, 
$(\alpha_{L,\text{QHA}}-\alpha_{L,\text{SCP}})/\alpha_{L,\text{QHA}} \sim 4\times {\braket{\hat{U}_4}}/{\braket{\hat{U}_2}}$, which is consistent with our estimation of Eq.~(\ref{eq_alpha_perturbation}).
Figure \ref{fig_MgO_Gamma_optical_freq_shift} is the $T$-dependent frequency shift of the optical modes of MgO at the $\Gamma$ point. 
The effect of the quartic anharmonicity in SCP is positive and cancels with $\Delta \Omega^{\text{Hes}}_{\bm{k}\lambda}$. This explains why the QHA works well for the phonon frequency shift in this material.

\section{Conclusion} 
We formulate a theory of thermal expansion based on the self-consistent phonon (SCP) theory to take into account the effect of the intrinsic lattice anharmonicity, i.e., phonon-phonon interaction, in a nonperturbative way. 
We compare the theory with QHA to derive the explicit formulae for the anharmonic corrections to the physical quantities regarding the thermal expansion.
We show that the Gr\"uneisen formula rigorously holds within the SCP theory by replacing the frequency in the original Gr\"uneisen theory with the SCP frequency. In addition to the QHA term, the phonon frequency shift have additional two contributions that come from $\Omega_{\bm{k}\lambda} - \omega_{\bm{k}\lambda}$ and $\Delta \Omega^{\text{Hes}}_{\bm{k}\lambda}$.
By performing perturbation expansion, we show that the QHA can calculate the thermal expansion coefficient with an accuracy of
$O(\braket{\hat{U}_4}/\braket{\hat{U}_2})$, while 
$| {\braket{\hat{U}_4}}/{\braket{\hat{U}_2}} |$ have to be much smaller than $10^{-3}\sim10^{-4}$ for $\Omega_{\bm{k}\lambda} - \omega_{\bm{k}\lambda}$ to be much smaller than the corresponding QHA term, which is too strict a condition for many existing materials.

Furthermore, we verified our theory by performing the first-principles calculations on silicon, diamond, NaCl, and MgO by utilizing the VASP and the ALAMODE package. We numerically showed that the two main anharmonic contributions of the phonon frequency shift $\Omega_{\bm{k}\lambda} - \omega_{\bm{k}\lambda}$ and $\Delta \Omega^{\text{Hes}}_{\bm{k}\lambda}$ are in the same order. When $\braket{\hat{U}_4}$ is positive, the two terms tend to accidentally cancel with each other, which can explain why the QHA correctly reproduces the $T$-dependent phonon frequency shift in some materials.
We infer that this cancellation occur in a wide range of anharmonic materials because $\braket{\hat{U}_4}$ is usually positive. On the other hand, this cancellation also shows that the QHA does not necessarily give reliable results for thermal expansion even when it appears to reproduce experimental $T$-dependence of phonon frequencies.
 We summarize the above discussion on the applicable limit of the QHA in Table \ref{table_summary_QHA_applicable_limit}

\begin{acknowledgments}
This work was supported by a Grant-in-Aid for Scientific  Research (No. 19H05825 and No. 21K03424), ``Program for Promoting Researches on the Supercomputer Fugaku'' (Project ID: hp200132) from MEXT.
\end{acknowledgments}

\appendix

\section{The derivation of the Gr\"uneisen formula for thermal expansion}
\label{Appendix_sec_Gruneisen_formula}
We review fhe derivation of the Gr\"uneisen formula~\cite{https://doi.org/10.1002/andp.19123441202} by following the derivation in Ref.~\cite{doi:10.1063/1.5125779}.

By using Maxwell's relation, the product of the volume thermal expansion coefficient $\alpha = \frac{1}{V}(\partial V/\partial T)_P$ and the bulk modulus $B_T = -V(\partial P/\partial V)_T$ can be written as
\begin{equation}
    \alpha B_T = \left( \frac{\partial P}{\partial T} \right)_V =  \left( \frac{\partial S}{\partial V} \right)_T .
    \label{eq_QHA_alpha_BT}
\end{equation}
In the Gr\"uneisen theory or QHA, we neglect all the anharmonic effect except for the $V$-dependence of the phonon frequency. Differentiating the QHA free energy
\begin{multline}
F_{\text{QHA}}(V,T) \\
= E_{\text{gnd}}(V) + \sum_{\bm{k}\lambda} \Bigl[ \frac{1}{2} \hbar \omega_{\bm{k}\lambda}(V) + k_{\mathrm{B}} T \log (1- e^{-\beta \hbar \omega_{\bm{k}\lambda}(V)}) \Bigr],
\label{Appendix_eq_QHA_free_energy}
\end{multline}
we get the entropy in this theory as 
\begin{multline}
S(T,V) = - \Bigl( \frac{\partial F}{\partial T} \Bigr)_V \\
= -\sum_{\bm{k}\lambda} \bigl[ k_{\mathrm{B}} \log (1 - e^{-\beta\hbar \omega_{\bm{k}\lambda}}) - \frac{\hbar \omega_{\bm{k}\lambda}}{T} n_B(\hbar \omega_{\bm{k}\lambda})\bigr].
  \label{eq_QHA_entropy}
\end{multline}
By substituting Eq. (\ref{eq_QHA_entropy}) to Eq. (\ref{eq_QHA_alpha_BT}), we get
\begin{equation}
\alpha^{\text{QHA}} = \frac{1}{B^{\text{QHA}}_T} \frac{1}{V}\sum_{\bm{k}\lambda} c^{\text{QHA}}_{v,\bm{k}\lambda} \gamma^{\text{QHA}}_{\bm{k}\lambda},
\end{equation}
where the mode specific heat $c^{\text{QHA}}_{v,\bm{k}\lambda}$ and the Gr\"uneisen parameter $\gamma^{\text{QHA}}_{\bm{k}\lambda}$ are defined as
\begin{align}
 c^{\text{QHA}}_{v,\bm{k}\lambda}(V,T) &=
  \frac{(\hbar \omega_{\bm{k}\lambda})^2}{k_{\mathrm{B}}T^2} n_B(\hbar \omega_{\bm{k}\lambda}) (n_B(\hbar \omega_{\bm{k}\lambda}) + 1) \notag \\
  &=
  \hbar  \omega_{\bm{k}\lambda} \frac{\partial  n_B(\hbar \omega_{\bm{k}\lambda} )}{\partial T}, \\
\gamma^{\text{QHA}}_{\bm{k}\lambda}(V) &= - \frac{V}{\omega_{\bm{k}\lambda}(V)} \Bigl( \frac{d \omega_{\bm{k}\lambda}(V)}{d V} \Bigr).
\end{align}

\section{The Taylor expansion of the potential energy surface (PES)}
\label{Appendix_sec_Taylor_expansion}
To formulate a theory that can incorporate the effects of lattice anharmonicity beyond QHA,
we start from the Taylor expansion of the PES:
\begin{equation}
\hat{U} = \sum_{n=0}^{\infty} \hat{U}_n,
\end{equation}
where
\begin{align}
&\hat{U}_n \nonumber \\
&=\frac{1}{n!} \sum_{\{\bm{R}\alpha \mu\}} \Phi_{\mu_1 \cdots \mu_n}(\bm{R}_1\alpha_1, \cdots, \bm{R}_n \alpha_n) \hat{u}_{\bm{R}_1 \alpha_1 \mu_1} \cdots \hat{u}_{\bm{R}_n \alpha_n \mu_n}
  \nonumber \\
&=\frac{1}{n!} \frac{1}{N^{n/2-1}} \sum_{\bm{k}\lambda} \delta_{\bm{k}_1 + \cdots + \bm{k}_n} \widetilde{\Phi} (\bm{k}_1 \lambda_1, \cdots, \bm{k}_n \lambda_n ) \hat{q}_{\bm{k_1} \lambda_1} \cdots \hat{q}_{\bm{k_n }\lambda_n}.
  \label{eq_Un}
\end{align}
Here, $\hat{u}_{\bm{R}\alpha\mu}$ is the $\mu(=x,y,z)$ component of the displacement of the atom $\alpha$ in the unit cell $\bm{R}$. The quantities defined as
\begin{equation}
\Phi_{\mu_1 \cdots \mu_n}(\bm{R}_1\alpha_1, \cdots, \bm{R}_n \alpha_n)
= \frac{\partial^n U}{\partial {u}_{\bm{R}_1 \alpha_1 \mu_1} \cdots \partial u_{\bm{R}_n \alpha_n \mu_n}}\Bigg|_{u=0}
\end{equation}
are called the $n$-th order interatomic force constants (IFCs) in real-space representation. In particular, the second-order IFCs are called the harmonic IFCs, the third-order IFCs as the cubic IFCs, and the fourth-order IFCs as the quartic IFCs. The Fourier transformation to the reciprocal space can be readily performed using the normal coordinate defined as
\begin{equation}
\hat{q}_{\bm{k}\lambda} = \frac{1}{N}\sum_{\bm{R}\alpha \mu} 
e^{-i\bm{k}\cdot \bm{R}} 
\epsilon^*_{\bm{k}\lambda,\alpha\mu}\sqrt{M_\alpha} \hat{u}_{\bm{R}\alpha\mu},
\end{equation}
where $M_{\alpha}$ is the atomic mass of atom $\alpha$, and $\epsilon_{\bm{k}\lambda,\alpha\mu}$ is the $\mu$ component of the polarization vector.
Then, the IFCs in the reciprocal-space ($k$-space) representation become
\begin{widetext}
\begin{align}
\widetilde{\Phi}(\bm{k}_1 \lambda_1, \cdots, \bm{k}_n \lambda_n ) 
&=\frac{1}{N}\sum_{\{\bm{R}\alpha\mu\}} \Phi_{\mu_1 \cdots \mu_n}(\bm{R}_1\alpha_1, \cdots, \bm{R}_n \alpha_n) 
\frac{\epsilon_{\bm{k}_1 \lambda_1,\alpha_1 \mu_1}}{\sqrt{M_{\alpha_1}}} e^{i\bm{k}_1\cdot \bm{R}_1} \cdots \frac{\epsilon_{\bm{k}_n \lambda_n,\alpha_n \mu_n}}{\sqrt{M_{\alpha_n}}} e^{i\bm{k}_n \cdot \bm{R}_n}
  \nonumber \\
&=\sum_{\{\alpha \mu\}} \frac{\epsilon_{\bm{k}_1 \lambda_1,\alpha_1 \mu_1}}{\sqrt{M_{\alpha_1}}}  \cdots \frac{\epsilon_{\bm{k}_n \lambda_n,\alpha_n \mu_n}}{\sqrt{M_{\alpha_n}}} 
  \sum_{\bm{R}_1\cdots \bm{R}_{n-1}} 
  \Phi_{\mu_1 \cdots \mu_n}(\bm{R}_1\alpha_1, \cdots, \bm{R}_{n-1} \alpha_{n-1}, \bm{0}\alpha_n)
  e^{i(\bm{k}_1\cdot \bm{R}_1+\cdots\bm{k}_{n-1} \cdot \bm{R}_{n-1})}.
  \nonumber
\end{align}
\end{widetext}
Note that $\lambda$ is the index of the phonon mode, which diagonalizes the harmonic part of the PES as
\begin{align}
&\sum_{\beta \nu} \frac{\Phi_{\mu\nu}(-\bm{k}\alpha, \bm{k}\beta)}{\sqrt{M_{\alpha} M_{\beta}}} \epsilon_{\bm{k}\lambda,\beta \nu} = \omega_{\bm{k}\lambda} \epsilon_{\bm{k}\lambda,\alpha \mu} ,
  \\
& \Phi_{\mu\nu}(-\bm{k}\alpha, \bm{k}\beta) =  \sum_{R} \Phi_{\mu\nu}(\bm{R}\alpha, \bm{0}\beta) e^{-i\bm{k} \cdot \bm{R}}.
\end{align}

\section{The self-consistent phonon(SCP) theory}
\label{Appendix_sec_SCP}
In this Appendix, we show the detailed calculation of the SCP equation and the Hessian of the SCP free energy.
The self-consistent phonon (SCP) theory is based on the variational principle of the free energy.
The \textit{effective} harmonic Hamiltonian
\begin{equation}
\hat{\mathcal{H}}_0 = \sum_{\bm{k}\lambda} \hbar \Omega_{\bm{k}\lambda} \Bigl(\hat{n}_{\bm{k}\lambda} + \frac{1}{2}\Bigr),
\end{equation}
is employed as the trial Hamiltonian,
where the frequencies $\Omega_{\bm{k}\lambda}$ are the variational parameters.
We assume that the change of the polarization vectors by the anharmonic renormalization can be neglected and use fixed mode approximation.
It should be noted that the definition of $\hat{n}_{\bm{k}\lambda}$ or $\hat{a}_{\bm{k}\lambda}$ and $\hat{a}^\dag_{\bm{k}\lambda}$ depend on the value of $\Omega_{\bm{k}\lambda}$.
The variational free energy is calculated analytically as 
\begin{align}
&
\mathcal{F}_1 (V, T, \Omega)
  = 
  \mathcal{F}_0 + \braket{\hat{H} - \hat{\mathcal{H}}_0}_{\hat{\mathcal{H}}_0}
  \nonumber
  \\&=
  U_0 + \sum_{\bm{k}\lambda} \Bigl[ \frac{1}{2} \hbar \Omega_{\bm{k}\lambda} + k_{\mathrm{B}} T \log (1- e^{-\beta \hbar \Omega_{\bm{k}\lambda}}) \Bigr] + \braket{\hat{H} - \hat{\mathcal{H}}_0}_{\hat{\mathcal{H}}_0}
  \nonumber
  \\&=
  \Phi_0 + 
  \Bigl[ \frac{1}{2} \hbar \Omega_{\bm{k}\lambda} + k_{\mathrm{B}} T \log (1- e^{-\beta \hbar \Omega_{\bm{k}\lambda}}) \Bigr]
  \nonumber
  \\&+\sum_{\bm{k}\lambda} (\omega_{\bm{k}\lambda}^2 - \Omega_{\bm{k}\lambda}^2)  g(\hbar \Omega_{\bm{k}\lambda})
  \nonumber
  \\&+
  \sum_{n=2}^{\infty} \frac{1}{n! N^{n-1}} \sum_{\bm{k}_1 \lambda_1, \cdots, n\lambda_n}
  \nonumber
  \widetilde{\Phi}(\bm{k}_1 \lambda_1, -\bm{k}_1 \lambda_1, \cdots, \bm{k}_n \lambda_n, -\bm{k}_n \lambda_n)
  \nonumber
  \\&\times 
  g(\hbar \Omega_{\bm{k}_1\lambda_1}) \cdots g(\hbar \Omega_{\bm{k}_n \lambda_n}),
  \label{Appendix_eq_SCP_free_energy_taylor_expand}
\end{align}
where $g(\Omega) = \frac{\hbar}{2\Omega} ( n_B(\hbar \Omega) + \frac{1}{2} )$.
By calculating the stationary condition with respect to the variational parameters, 
we get the SCP equation 
\begin{align}
\Omega_{\bm{k}\lambda}^2 =&
\omega_{\bm{k}\lambda}^2 
  \nonumber
  +
  \sum_{n=2}^\infty \frac{1}{(n-1)! N^{n-1}} \sum_{\bm{k}_1\lambda_1, \cdots, \bm{k}_{n-1}\lambda_{n-1}} 
  \nonumber
  \\&
  \times
  \widetilde{\Phi}(\bm{k} \lambda, -\bm{k} \lambda, \bm{k}_1 \lambda_1, -\bm{k}_1 \lambda_1, \cdots, \bm{k}_{n-1} \lambda_{n-1}, -\bm{k}_{n-1} \lambda_{n-1})
  \nonumber
  \\&\times 
  g(\Omega_{\bm{k}_1\lambda_1}) \cdots g(\Omega_{\bm{k}_{n-1}\lambda_{n-1}}),
  \label{Appendix_eq_SCP_equation}
\end{align}

Here, we move on to the calculation of the Hessian of the SCP free energy. We calculate the Hessian of the SCP free energy because it is interpreted as the renormalized anharmonic phonon frequency~\cite{PhysRevB.96.014111}, instead of the SCP frequency $\Omega_{\bm{k}\lambda}$.
We consider the diagonal part of the Hessian because we use the fixed-mode approximation.
When a static atomic displacement is introduced in the system, the expectation values of the normal coordinate operators $\hat{q}_{\bm{k}\lambda}$ become finite; we denote this as $q_{\bm{k}\lambda}$, without the hat on $q$. 
Taylor expanding the PES with respect to the atomic displacements from the new static positions corresponds to replacing the $\hat{q}_{\bm{k}\lambda}$ in Eq. (\ref{eq_Un}) by $q_{\bm{k}\lambda} + \hat{q}_{\bm{k}\lambda}$. 
The SCP free energy $\mathcal{F}_1$ is a function of $V$, $T$, and $q = \{q_{\bm{k}\lambda}\}$, where $\mathcal{F}_1$ is calculated by using the SCP frequency that satisfies the SCP equation at each point in the $(V, T, q)$ space. 

We calculate the Hessian for the $\bm{k}=0$ case.
The final result, which is equivalent to the SCP+QP[0] theory~\cite{tadano2021firstprinciples} implemented in the ALAMODE package, is shown in Eq. (\ref{eq_QP0_squared_freq}).
The result can be extended to finite $\bm{k}$ by considering a commensurate supercell.
\begin{align}
&
  \frac{\partial^2 \mathcal{F}_1(V,T,q)}{\partial q_{\bm{0}\lambda}^{2}}
  \nonumber
  \\
  =&
  \Bigl( \frac{\partial}{\partial q_{\bm{0}\lambda}} + 
  \sum_{\bm{k}\lambda_3 \lambda_4} 
  \frac{\partial (\Omega^2_{\bm{k}\lambda_3 \lambda_4}) }{\partial q_{\bm{0}\lambda}} 
  \frac{\partial}{\partial ( \Omega^2_{\bm{k}\lambda_3 \lambda_4} ) }\Bigr)
  \nonumber
  \\&
  \Bigl( \frac{\partial}{\partial q_{\bm{0}\lambda}} + 
  \sum_{\bm{k}\lambda_1 \lambda_2} 
  \frac{\partial ( \Omega^2_{\bm{k}\lambda_1 \lambda_2}) }{\partial q_{\bm{0}\lambda}} 
  \frac{\partial}{\partial (\Omega^2_{\bm{k}\lambda_1 \lambda_2})}\Bigr) 
  \mathcal{F}_1(V,T,q,\Omega)
  \label{eq_SCP_Hessian1}
\end{align}
where $\Omega^2_{\bm{k}\lambda_1\lambda_2}$ are the components of the $\Omega^2$ matrix, which extends the SCP frequency to include the off-diagonal terms, whose dominant contribution is given by $\widetilde{\Phi}(-\bm{k}\lambda_1, \bm{k}\lambda_2)$.
The off-diagonal terms appear because the atomic displacements change the harmonic IFC and the polarization vector.
Because $\Bigl( \frac{\partial}{\partial q_{\bm{0}\lambda}} + 
  \sum_{\bm{k}\lambda_3 \lambda_4} 
  \frac{\partial (\Omega^2_{\bm{k}\lambda_3 \lambda_4}) }{\partial q_{\bm{0}\lambda}} 
\frac{\partial}{\partial ( \Omega^2_{\bm{k}\lambda_3 \lambda_4} ) }\Bigr)$ is a derivative along the solution of the SCP equation in the $(V, T, q, \Omega)$ space, the variational condition is maintained along this direction. Therefore,
\begin{align}
&
\sum_{\bm{k}\lambda_1 \lambda_2} 
\frac{\partial ( \Omega^2_{\bm{k}\lambda_1 \lambda_2}) }{\partial q_{\bm{0}\lambda}} 
  \Bigl( \frac{\partial}{\partial q_{\bm{0}\lambda}} + 
  \sum_{\bm{k}\lambda_3 \lambda_4} 
  \frac{\partial (\Omega^2_{\bm{k}\lambda_3 \lambda_4}) }{\partial q_{\bm{0}\lambda}} 
  \frac{\partial}{\partial ( \Omega^2_{\bm{k}\lambda_3 \lambda_4} ) }\Bigr)
  \nonumber
  \\&\times 
  \frac{\partial\mathcal{F}_1(V,T,q,\Omega)}{\partial (\Omega^2_{\bm{k}\lambda_1 \lambda_2})} 
  \\&
  \simeq 0,
\end{align}
where we used that the diagonal part of $  \frac{\partial\mathcal{F}_1(V,T,q,\Omega)}{\partial (\Omega^2_{\bm{k}\lambda_1 \lambda_2})} 
$ vanishes due to the variational condition. Thus, we get
\begin{align}
&
  \frac{\partial^2 \mathcal{F}_1(V,T,q)}{\partial q_{\bm{0}\lambda}^2}
  \nonumber
  \\&
  \simeq
  \frac{\partial^2 \mathcal{F}_1(V,T,q,\Omega)}{\partial q_{\bm{0}\lambda}^2} 
  + 
  \sum_{\bm{k}\lambda_3 \lambda_4} 
  \frac{\partial (\Omega^2_{\bm{k}\lambda_3 \lambda_4}) }{\partial q_{\bm{0}\lambda}} 
  \frac{\partial^2 \mathcal{F}_1(V,T,q,\Omega)}{\partial q_{\bm{0}\lambda} \partial (\Omega^2_{\bm{k}\lambda_3 \lambda_4} ) }
  \label{eq_approcimate_Hessian1}
\end{align}
The first term of the RHS of Eq. (\ref{eq_approcimate_Hessian1}) is calculated as
\begin{align}
  &\frac{\partial^2 \mathcal{F}_1(V,T,q, \Omega)}{\partial q_{\bm{0}\lambda}^2}
  \nonumber
  \\=&
  \omega^2_{\bm{0}\lambda} 
  + 
  \sum_{n = 1}^{\infty} \frac{1}{n!} \frac{1}{N^{n}} 
  \sum_{\bm{k}_1\lambda_1, \cdots, \bm{k}_n\lambda_n}
  \nonumber
  \\&\times
  \widetilde{\Phi}(-\bm{0}\lambda,\bm{0}\lambda, \bm{k}_1 \lambda_1, -\bm{k}_1 \lambda_1, \cdots, -\bm{k}_n \lambda_n)
  \nonumber
  \\&
  \times 
  g(\Omega_{\bm{k}_1 \lambda_1})
  \cdots 
  g(\Omega_{\bm{k}_n \lambda_n})
  \nonumber
  \\=&
  \Omega_{\bm{0}\lambda}^2,
\end{align}
by using the SCP equation. 

Because the dominant contribution to $\Omega^2_{\bm{k}\lambda_1 \lambda_2}$ is $\widetilde{\Phi}(-\bm{k}\lambda_1, \bm{k}\lambda_2)$, its derivative is approximated as
\begin{align}
\frac{\partial (\Omega^2_{\bm{k}_1\lambda_3 \lambda_4}) }{\partial q_{\bm{0}\lambda}} \simeq \frac{1}{\sqrt{N}}\widetilde{\Phi}(-\bm{k}_1\lambda_3, \bm{k}_1\lambda_4, \bm{0} \lambda)
\end{align}
Thus, we get
\begin{align}
&
  \frac{\partial^2 \mathcal{F}_1}{ \partial q_{\bm{0}\lambda} \partial (\Omega^2_{\bm{k}\lambda_3 \lambda_4} ) }
  \nonumber
  \\=&
  \frac{\partial}{\partial (\Omega^2_{\bm{k}\lambda_3 \lambda_4} )}
  \frac{\partial}{\partial q_{\bm{0}\lambda}} \sum_{\bm{k}_1 \lambda' \lambda_1 \lambda_2}
  g(\Omega_{\bm{k}_1 \lambda'})
  \nonumber
  \\&\times
  C_{\bm{k}_1 \lambda_1 \lambda'}^* \widetilde{\Phi}(-\bm{k}_1 \lambda_1, \bm{k}_1 \lambda_2) C_{\bm{k}_1 \lambda_2 \lambda'}
  \nonumber
  \\=&
  \frac{\partial}{\partial (\Omega^2_{\bm{k}\lambda_3 \lambda_4} )}
  \sum_{\bm{k}_1 \lambda' \lambda_1 \lambda_2}
  g(\Omega_{\bm{k}_1 \lambda'})
  \nonumber
  \\&
  \frac{1}{\sqrt{N}}
  C_{\bm{k}_1 \lambda_1 \lambda'}^* \widetilde{\Phi}(-\bm{k}_1 \lambda_1, \bm{k}_1 \lambda_2, \bm{0}\lambda) C_{\bm{k}_1 \lambda_2 \lambda'}
  \nonumber
  \\=&
  \frac{1}{\sqrt{N}}
  \sum_{ \lambda' \lambda_1 \lambda_2}
  \delta_{\lambda_3 \lambda'} \delta_{\lambda_4 \lambda'} \frac{\hbar}{2} 
  \frac{\partial}{\partial(\Omega^2_{\bm{k}\lambda'})} \Bigl( \frac{n_B(\hbar \Omega_{\bm{k}\lambda'}) + 1/2}{\Omega_{\bm{k}\lambda'}} \Bigr)
  \nonumber
  \\&\times
  C_{\bm{k}_1 \lambda_1 \lambda'}^* \widetilde{\Phi}(-\bm{k}_1 \lambda_1, \bm{k}_1 \lambda_2, \bm{0}\lambda) C_{\bm{k}_1 \lambda_2 \lambda'}
  \nonumber
  \\&+
  \frac{1}{\sqrt{N}}
  \sum_{ \lambda' \lambda_1 \lambda_2} 
  g(\Omega_{\bm{k}\lambda'})
  \frac{\partial C_{\bm{k}_1 \lambda_1 \lambda'}^*}{\partial  (\Omega^2_{\bm{k}\lambda_3 \lambda_4} )} \widetilde{\Phi}(-\bm{k}_1 \lambda_1, \bm{k}_1 \lambda_2, \bm{0}\lambda) C_{\bm{k}_1 \lambda_2 \lambda'}
  \nonumber
  \\&+
  \frac{1}{\sqrt{N}}
  \sum_{ \lambda' \lambda_1 \lambda_2}
  g(\Omega_{\bm{k}\lambda'})
  C_{\bm{k}_1 \lambda_1 \lambda'}^* \widetilde{\Phi}(-\bm{k}_1 \lambda_1, \bm{k}_1 \lambda_2, \bm{0}\lambda) \frac{\partial C_{\bm{k}_1 \lambda_2 \lambda'}}{\partial (\Omega^2_{\bm{k}\lambda_3 \lambda_4} )},
  \label{Appendix_eq_partial2F_partialq_partialomega}
\end{align}
where $C_{\bm{k}\lambda \lambda'} = \sum_{\alpha\mu} \epsilon^*_{\bm{k}\lambda,\alpha \mu} \epsilon_{\bm{k}\lambda',\alpha \mu}$ describes the change of polarization vector induced by the static atomic displacements. The $\lambda$(without prime) denotes the modes with the fixed polarization vector, and the $\lambda'$(with prime) denotes the new polarization vector which is changed by the atomic displacements. Note that the summations in the last two terms in the RHS of Eq. (\ref{Appendix_eq_partial2F_partialq_partialomega}) is taken for modes that satisfy $\Omega_{\bm{k}\lambda_1}\neq\Omega_{\bm{k}\lambda_2} $ because it is possible to eliminate the contribution from the degenerate modes due to the freedom in the choice of polarization vectors.
Because $\epsilon_{\bm{k}\lambda'}$ diagonalizes $\Omega^2_{\bm{k}\lambda_1 \lambda_2}$, we can show that
\begin{align}
  \frac{\partial C^*_{\bm{k}\lambda_1 \lambda'}}{\partial (\Omega^2_{\bm{k}\lambda_3 \lambda_4})}
  &=
  \sum_{{\lambda'}_1} \frac{ C^*_{\bm{k}\lambda_3 \lambda'} C_{\bm{k}\lambda_4 {\lambda_1}'}} {\Omega^2_{\bm{k}\lambda'} - \Omega^2_{\bm{k}\lambda'_1}}
  C^*_{\bm{k}\lambda_1 \lambda'_1}
  \nonumber
  \\&=
  \frac{ \delta_{\lambda_3 \lambda'} \delta_{\lambda_4 {\lambda_1}}} {\Omega^2_{\bm{k}\lambda'} - \Omega^2_{\bm{k}\lambda_4}}
  \label{Appendix_eq_Cdag_Omega2} \\
  \frac{\partial C_{\bm{k}\lambda_2 \lambda'}}{\partial (\Omega^2_{\bm{k}\lambda_3 \lambda_4})}
  &=
  \sum_{{\lambda'}_1} C_{\bm{k}\lambda_2\lambda'_1} \frac{  C^*_{\bm{k}\lambda_3 {\lambda_1}'} C_{\bm{k}\lambda_4 \lambda'}} {\Omega^2_{\bm{k}\lambda'} - \Omega^2_{\bm{k}\lambda'_1}}
  \nonumber
  \\&=
  \frac{ \delta_{\lambda_2 \lambda_3} \delta_{\bm{k}\lambda_4 {\lambda'}}} {\Omega^2_{\bm{k}\lambda'} - \Omega^2_{\bm{k}\lambda_2}}.
  \label{Appendix_eq_C_Omega2}
\end{align}
Note that $C_{\bm{k}\lambda \lambda'} = \delta_{\lambda \lambda'}$ at $q=0$, which we are considering.
Substituting Eq. (\ref{Appendix_eq_partial2F_partialq_partialomega}), (\ref{Appendix_eq_Cdag_Omega2}), and (\ref{Appendix_eq_C_Omega2}) to Eq. (\ref{eq_approcimate_Hessian1}), we get
\begin{align}
  \frac{\partial^2 \mathcal{F}_1}{\partial q^{2}_{\bm{0}\lambda}}
  \nonumber\
  =&
  \Omega_{\bm{0}\lambda}^2
  \nonumber
  \\&+
  \sum_{\bm{k}\lambda_1} \frac{\hbar}{4} \frac{|\widetilde{\Phi}(-\bm{k}\lambda_1, \bm{k}\lambda_1, \bm{0}\lambda)|^2}{\Omega_{\bm{k}\lambda_1^2}} 
  \nonumber
  \\&\times
  \Bigl( \frac{\partial n_B(\hbar \Omega_{\bm{k}\lambda_1})}{\Omega_{\bm{k}\lambda_1}}  - \frac{2n_B(\hbar \Omega_{\bm{k}\lambda_1}) + 1}{2\Omega_{\bm{k}\lambda_1}}\Bigr)
  \nonumber
  \\&+
  \sum_{\bm{k}\lambda_1 \lambda_2(\Omega_{\bm{k}\lambda_1} \neq \Omega_{\bm{k}\lambda_2} )} 
  \frac{\hbar}{4} \frac{| \widetilde{\Phi}(-\bm{k}\lambda_1, \bm{k}\lambda_2, \bm{0}\lambda) |^2}{ \Omega_{\bm{k}\lambda_1} \Omega_{\bm{k}\lambda_2} }
  \nonumber
  \\&\times
  \Bigl(
    \frac{n_B(\hbar \Omega_{\bm{k}\lambda_1}) - n_B(\hbar \Omega_{\bm{k}\lambda_2})}{\Omega_{\bm{k}\lambda_1} - \Omega_{\bm{k}\lambda_2}}
    \nonumber
    \\&
    - 
    \frac{n_B(\hbar \Omega_{\bm{k}\lambda_1}) + n_B(\hbar \Omega_{\bm{k}\lambda_2})+1}{\Omega_{\bm{k}\lambda_1} + \Omega_{\bm{k}\lambda_2}}
  \Bigr)
  \label{Appendix_eq_QP0_squared_freq}
\end{align}

\section{The Derivation of the SCP entropy}
\label{Appendix_sec_SCP_entropy}
The SCP entropy is calculated by differentiating the SCP free energy $\mathcal{F}_1$. Because the final form of the SCP entropy has been derived in the previous research that use different formalisms of SCP~\cite{PhysRevB.1.572, Hui_1975}, we show another derivation that uses the direct expansion to infinite orders.
Using the variational condition with respect to $\Omega_{\bm{k}\lambda}$,
\begin{align}
&S(V,T)
  =
  - \Bigl( \frac{\partial \mathcal{F}_1 (V, T)}{\partial T} \Bigr)_V
  \nonumber
  \\&=
  - \Bigl( \frac{\partial \mathcal{F}_1(V,T, \Omega)}{\partial T} \Bigr)_{V, \Omega} 
  - \sum_{\bm{k}\lambda} \Bigl( \frac{\partial \Omega_{\bm{k}\lambda}}{\partial T} \Bigr)_V \Bigl( \frac{\partial \mathcal{F}_1(V,T, \Omega)}{\partial \Omega_{\bm{k}\lambda}} \Bigr)_{V, T} 
  \nonumber
  \\&=
  - \Bigl( \frac{\partial \mathcal{F}_1(V, T, \Omega)}{\partial T} \Bigr)_{V, \Omega}
\end{align}
By differentiating Eq. (\ref{Appendix_eq_SCP_free_energy_taylor_expand}) and using the SCP equation, we get
\begin{align}
&
S(V,T)
\nonumber
\\=&
  -\sum_{\bm{k}\lambda} \bigl[ k_{\mathrm{B}} \log (1 - e^{-\beta\hbar \Omega_{\bm{k}\lambda}}) - \frac{\hbar \Omega_{\bm{k}\lambda}}{T} n_B(\hbar \Omega_{\bm{k}\lambda})\bigr]
  \nonumber
  \\&
  - \sum_{\bm{k}\lambda} \Bigl[ \frac{\hbar \Omega_{\bm{k}\lambda}}{2}\Bigl( \frac{\partial n_B(\hbar \Omega_{\bm{k}\lambda})}{\partial T} \Bigr)_{\Omega} 
  \nonumber
  \\&\times 
  \Bigl\{ 
    \omega_{\bm{k}\lambda}^2 -\Omega_{\bm{k}\lambda}^2 
    + 
    \sum_{n=2}^\infty \frac{1}{(n-1)! N^{n-1}} 
    \sum_{\bm{k}_1\lambda_1\cdots \bm{k}_{n-1}\lambda_{n-1}}
     \nonumber
    \\&\times
    \widetilde{\Phi}(\bm{k} \lambda, -\bm{k} \lambda, \bm{k}_1 \lambda_1, -\bm{k}_1 \lambda_1, \cdots, \bm{k}_n \lambda_n, -\bm{k}_n \lambda_n)
    \nonumber
    \\&\times 
    g(\Omega_{\bm{k}_1\lambda_1}) \cdots g(\Omega_{\bm{k}_{n-1}\lambda_{n-1}})
  \Bigr\}
  \Bigr]
  \nonumber
  \\=&
  -\sum_{\bm{k}\lambda} \bigl[ k_{\mathrm{B}} \log (1 - e^{-\beta\hbar \Omega_{\bm{k}\lambda}}) - \frac{\hbar \Omega_{\bm{k}\lambda}}{T} n_B(\hbar \Omega_{\bm{k}\lambda})\bigr].
\end{align}
It should be noted that the SCP entropy has the same form as the entropy of the harmonic Hamiltonian except that the phonon frequency is replaced by $\Omega_{\bm{k}\lambda}$.
\section{The perturbation expansion}
In this Appendix, we show the details of the derivations or the calculations that are skipped in section \ref{sec_perturbation_expansion}.

\subsection{The SCP frequency}

\label{Appendix_subsec_SCP_freq}
In section \ref{subsec_SCP_freq}, we showed that the lowest-order estimate of $\Omega_{\bm{k}\lambda} - \omega_{\bm{k}\lambda}$ is given by 
\begin{equation}
  \Omega_{\bm{k}\lambda} - \omega_{\bm{k}\lambda} \simeq \omega_{\bm{k}\lambda} \times \frac{\braket{\hat{U}_4}}{\braket{\hat{U}_2}}.
\end{equation}
In this Appendix, we consider the higher-order terms of the expansion of $\Omega_{\bm{k}\lambda} - \omega_{\bm{k}\lambda}$.
The next dominant contribution can be the following two terms. The first one is the correction by that we displaced $\Omega_{\bm{k}\lambda}$ in the RHS of the SCP equation by $\Omega_{\bm{k}\lambda}$ in the estimation of Eq. (\ref{eq_Omega_omega_lowest_order1}).
\begin{align}
  &
  \Delta_1(\Omega_{\bm{k}\lambda} - \omega_{\bm{k}\lambda} )
  \nonumber
  \\=&
  \frac{1}{N} \sum_{\bm{k}'\lambda'}
  \frac{
  \widetilde{\Phi}(\bm{k}\lambda,-\bm{k}\lambda, \bm{k}'\lambda',-\bm{k}'\lambda')
  }{(\Omega_{\bm{k}\lambda}+\omega_{\bm{k}\lambda}) 
  }
  g(\Omega_{\bm{k}'\lambda'})
  \nonumber
  \\&
  -
  \frac{1}{N} \sum_{\bm{k}'\lambda'} 
  \frac{1}{2}
  \frac{\widetilde{\Phi}(\bm{k}\lambda,-\bm{k}\lambda, \bm{k}'\lambda',-\bm{k}'\lambda')}{\omega_{\bm{k}\lambda} 
  }
  g(\omega_{\bm{k}'\lambda'})
  \nonumber
  \\\sim&
  \omega_{\bm{k}\lambda} \times O\Bigl( \Bigl( \frac{\braket{\hat{U}_4}}{\braket{\hat{U}_2}} \Bigr)^2 \Bigr).
\end{align}
The second one is the next lowest-order term of the RHS of the SCP equation:
\begin{align}
&
  \Delta_2(\Omega_{\bm{k}\lambda} - \omega_{\bm{k}\lambda} )
  \nonumber
  \\=&
  \frac{1}{2} \frac{1}{2N^2} \sum_{\bm{k}_1\lambda_1, \bm{k}_{2}\lambda_{2}} 
  \frac{\widetilde{\Phi}(\bm{k} \lambda, -\bm{k} \lambda, \bm{k}_1 \lambda_1, -\bm{k}_1 \lambda_1,\bm{k}_2 \lambda_2, -\bm{k}_2 \lambda_2)}{\omega_{\bm{k}\lambda}
  }
  \nonumber
  \\&\times 
  g(\omega_{\bm{k}_1\lambda_1}) g(\omega_{\bm{k}_{2}\lambda_{2}}).
\end{align}
By using
\begin{align}
  \braket{\hat{U}_6}
   \simeq&
  \frac{1}{6 N^{2}} \sum_{\{\bm{k} \lambda \}}
  \widetilde{\Phi}(\bm{k}_1 \lambda_1, -\bm{k}_1 \lambda_1, \cdots, \bm{k}_3 \lambda_3, -\bm{k}_3 \lambda_3)
  \nonumber
  \\&\times 
  g(\omega_{\bm{k}_1\lambda_1}) g(\omega_{\bm{k}_2\lambda_2}) g(\omega_{\bm{k}_3\lambda_3})
  \nonumber
  \\=&
  \sum_{\bm{k}\lambda} \Bigl[ \frac{1}{2} \Bigl( n_B(\hbar \omega_{\bm{k} \lambda}) + \frac{1}{2} \Bigr) 
  \times 
  \frac{\hbar}{6N^2} \sum_{\bm{k}_1\lambda_1,\bm{k}_2\lambda_2} 
  \nonumber
  \\&\times
  \frac{\widetilde{\Phi}(\bm{k} \lambda, -\bm{k} \lambda, \bm{k}_1 \lambda_1, -\bm{k}_1 \lambda_1, \bm{k}_2 \lambda_2, -\bm{k}_2 \lambda_2)}{\omega_{\bm{k} \lambda}
  }
  \nonumber
  \\&
  \times g(\omega_{\bm{k}_1\lambda_1}) g(\omega_{\bm{k}_2\lambda_2}),
\end{align}
we get the average estimation of this $\Delta_2$ as
\begin{align}
  \Delta_2(\Omega_{\bm{k}\lambda} - \omega_{\bm{k}\lambda}) \simeq \omega_{\bm{k}\lambda} \times \frac{3\braket{\hat{U}_6}}{2\braket{\hat{U}_2}}.
\end{align}
To summarize the above discussion, we get
\begin{align}
  \Omega_{\bm{k}\lambda} - \omega_{\bm{k}\lambda} = \omega_{\bm{k}\lambda} \times \Bigl[ \frac{\braket{\hat{U}_4}}{\braket{\hat{U}_2}} + O\Bigl( \frac{\braket{\hat{U}_4}^2}{\braket{\hat{U}_2}^2}\Bigr) + \frac{3\braket{\hat{U}_6}}{2\braket{\hat{U}_2}} + \cdots \Bigr]
  \nonumber
\end{align}
The higher-order expansion can be performed in a similar way, but the result will get extremely complicated as we go to the higher-orders.

\subsection{The Gr\"uneisen parameter, the mode specific heat, the bulk modulus and the thermal expansion coefficient}
\label{Appendix_subsec_Gruneisen_param_alpha}
Differentiating the SCP equation by the system volume, we get
\begin{align}
  &
  2\Omega_{\bm{k}\lambda} \frac{\partial \Omega_{\bm{k}\lambda}}{\partial V} - 2\omega_{\bm{k}\lambda} \frac{\partial \omega_{\bm{k}\lambda}}{\partial V}
  \nonumber
  \\=&
  \sum_{n=2}^{\infty} \frac{1}{(n-1)! N^{n-1}} 
  \sum_{\bm{k}_1\lambda_1\cdots \bm{k}_{n-1} \lambda_{n-1}} 
  \nonumber 
  \\&\times
  \frac{\partial \widetilde{\Phi}(\bm{k}\lambda,-\bm{k}\lambda, \bm{k}_1\lambda_1,-\bm{k}_1\lambda_1, \cdots, \bm{k}_{n-1}\lambda_{n-1},-\bm{k}_{n-1}\lambda_{n-1})}{\partial V}
  \nonumber
  \\&\times 
  g(\Omega_{\bm{k}_1\lambda_1}) \cdots g(\Omega_{\bm{k}_{n-1}\lambda_{n-1}})
  \nonumber
  \\&+
  \sum_{n=2}^{\infty} \frac{1}{(n-2)! N^{n-1}} 
    \sum_{\bm{k}_1\lambda_1\cdots \bm{k}_{n-1} \lambda_{n-1}} 
  \nonumber 
  \\&\times
  \widetilde{\Phi}(\bm{k}\lambda,-\bm{k}\lambda, \bm{k}_1\lambda_1,-\bm{k}_1\lambda_1, \cdots, \bm{k}_{n-1}\lambda_{n-1},-\bm{k}_{n-1}\lambda_{n-1})
  \nonumber
  \\&\times 
  g(\Omega_{\bm{k}_2\lambda_2}) \cdots g(\Omega_{\bm{k}_{n-1}\lambda_{n-1}})
  \nonumber
  \\&\times
  \frac{\partial}{\partial \Omega_{\bm{k}_1 \lambda_1}} \Bigl[ \frac{\hbar}{2 \Omega_{\bm{k}_1 \lambda_1}} \Bigl( n_B(\hbar \Omega_{\bm{k}_{1}\lambda_{1}}) + \frac{1}{2} \Bigr) \Bigr]
  \frac{\partial \Omega_{\bm{k}_1 \lambda_1}}{\partial V}.
\end{align}
Thus, the dominant correction to the Gr\"uneisen parameter is
\begin{align}
\gamma_{\bm{k}\lambda} 
  \simeq&
  \frac{\omega_{\bm{k}\lambda}^2}{\Omega_{\bm{k}\lambda}^2} \gamma^{\text{QHA}}_{\bm{k}\lambda} \notag \\
  &- 
  \frac{V}{2\Omega_{\bm{k}\lambda}^2} \frac{1}{N} \sum_{\bm{k}'\lambda'} 
  \frac{\partial \widetilde{\Phi}(\bm{k}\lambda,-\bm{k}\lambda,\bm{k}'\lambda',-\bm{k}'\lambda')}{\partial V}
  g(\Omega_{\bm{k}'\lambda'})
  \nonumber
  \\&
  - \frac{V}{2\Omega_{\bm{k}\lambda}^2} \sum_{\bm{k}'\lambda'} \frac{\hbar}{2} \widetilde{\Phi}(\bm{k}\lambda,-\bm{k}\lambda,\bm{k}'\lambda',-\bm{k}'\lambda')
  \nonumber
  \\&\times
  \frac{\partial }{\partial \Omega_{\bm{k}'\lambda'}}\Bigl[ \frac{1}{\Omega_{\bm{k}' \lambda'}} \Bigl( n_B(\hbar \Omega_{\bm{k}'\lambda'}) + \frac{1}{2} \Bigr) \Bigr]
  \frac{\partial \Omega_{\bm{k}' \lambda'}}{\partial V}
  \nonumber
  \\=&
  \gamma^{\text{QHA}}_{\bm{k}\lambda} + \Delta_1 \gamma_{\bm{k}\lambda} + \Delta_2 \gamma_{\bm{k}\lambda} + \Delta_3 \gamma_{\bm{k}\lambda},
  \label{Appendix_eq_delta123_gamma}
\end{align}
where $\gamma^{\text{QHA}}_{\bm{k}\lambda}$ is the Gr\"uneisen parameter calculated in the QHA.
We investigate each term in the RHS of Eq. (\ref{Appendix_eq_delta123_gamma}).
\begin{align}
  \gamma^{\text{QHA}}_{\bm{k}\lambda} + \Delta_1 \gamma_{\bm{k}\lambda} &=
  \frac{\omega_{\bm{k}\lambda}^2}{\Omega_{\bm{k}\lambda}^2} \gamma^{\text{QHA}}_{\bm{k}\lambda} 
  \nonumber
  \\&\simeq
  \Bigl( 1 - \frac{2\braket{\hat{U}_4}}{\braket{\hat{U}_2}} \Bigr) \gamma^{\text{QHA}}_{\bm{k}\lambda}
\end{align}
To get an estimation for the $\Delta_2 \gamma_{\bm{k}\lambda}$, we calculate the SCP pressure from the SCP free energy.
\begin{align}
  P =&
  - \Bigl( \frac{\partial \mathcal{F}_1(T,V)}{\partial V} \Bigr)_T
  \nonumber
  \\=&
  - \Bigl[ \Bigl( \frac{\partial}{\partial V} \Bigr)_{T, \Omega} 
  + 
  \sum_{\bm{k}\lambda} \Bigl( \frac{\partial \Omega_{\bm{k}\lambda}}{\partial V} \Bigr)_T \Bigl( \frac{\partial }{\partial \Omega_{\bm{k}\lambda}} \Bigr)_{V,T} \Bigr] 
  \mathcal{F}_1(T, V, \Omega)
  \nonumber
  \\=&
  \Bigl( \frac{\partial \mathcal{F}_1(T,V,\Omega)}{\partial V} \Bigr)_{T, \Omega} \text{\ \ (variational principle)}
  \nonumber
  \\
  =&
  -\frac{\partial \Phi_0}{\partial V}
  - \sum_{\bm{k}\lambda} 
  2\omega_{\bm{k}\lambda} \Bigl( \frac{\partial \omega_{\bm{k}\lambda}}{\partial V} \Bigr) 
  g(\Omega_{\bm{k}\lambda})
  \nonumber
  \\&-
  \sum_{n=2}^{\infty} \frac{1}{n!N^{n-1}} \sum_{\bm{k}_1 \lambda_1, \cdots, \bm{k}_{n}\lambda_{n}} 
  \nonumber
  \\&\times
  \frac{\partial \widetilde{\Phi}(\bm{k}_1\lambda_1,-\bm{k}_1\lambda_1, \cdots, \bm{k}_n\lambda_n,-\bm{k}_n\lambda_n)}{\partial V}
  \nonumber
  \\&\times
  g(\Omega_{\bm{k}_1\lambda_1}) \cdots g(\Omega_{\bm{k}_n\lambda_n}).
  \label{Appendix_eq_SCP_pressure}
\end{align}
This formula is equivalent to the previous result that use another formalism of SCP theory~\cite{PhysRevB.98.024106}.
We define
\begin{equation}
  P_2 = \frac{1}{2} \times \sum_{\bm{k}\lambda} - \hbar \frac{\partial \omega_{\bm{k}\lambda}}{\partial V} \Bigl( n_B(\hbar \Omega_{\bm{k}\lambda}) + \frac{1}{2} \Bigr)
\end{equation}
\begin{align}
  P_4 
  \simeq&
  \frac{-1}{2N} \sum_{\bm{k}\lambda,\bm{k}'\lambda'} 
  \frac{\partial \widetilde{\Phi}(\bm{k}\lambda,-\bm{k}\lambda,\bm{k}'\lambda',-\bm{k}'\lambda')}{\partial V}
  g(\Omega_{\bm{k}\lambda}) g(\Omega_{\bm{k}'\lambda'})
\end{align}
from the RHS of Eq. (\ref{Appendix_eq_SCP_pressure}). Note that $P_2$ is the half of the second term in the RHS of Eq. (\ref{Appendix_eq_SCP_pressure}) because the second term includes the contributions from the kinetic energy and the harmonic term of the potential. 
Thus, the average estimation of $\Delta_2 \gamma_{\bm{k}\lambda}$ is
\begin{equation}
  \Delta_2 \gamma_{\bm{k}\lambda} \simeq \gamma^{\text{QHA}}_{\bm{k}\lambda} \times \frac{P_4}{P_2}.
\end{equation}
Next, we continue with $\Delta_3 \gamma_{\bm{k}\lambda}$.
\begin{align}
&
\Delta_3 \gamma_{\bm{k}\lambda}
\nonumber
  \\=&
  - \frac{V}{2\Omega_{\bm{k}\lambda}^2} \sum_{\bm{k}'\lambda'} \frac{\hbar}{2} \widetilde{\Phi}(\bm{k}\lambda,-\bm{k}\lambda,\bm{k}'\lambda',-\bm{k}'\lambda')
  \nonumber
  \\&\times
  \frac{\partial }{\partial \Omega_{\bm{k}'\lambda'}}\Bigl[ \frac{1}{\Omega_{\bm{k}' \lambda'}} \Bigl( n_B(\hbar \Omega_{\bm{k}'\lambda'}) + \frac{1}{2} \Bigr) \Bigr]
  \frac{\partial \Omega_{\bm{k}' \lambda'}}{\partial V}
  \\\simeq&
  - \frac{V}{2\omega_{\bm{k}\lambda}} \sum_{\bm{k}'\lambda'} \frac{\hbar}{2} \widetilde{\Phi}(\bm{k}\lambda,-\bm{k}\lambda,\bm{k}'\lambda',-\bm{k}'\lambda')
  \nonumber
  \\&\times
  \frac{\partial }{\partial \omega_{\bm{k}'\lambda'}}\Bigl[ \frac{1}{\omega_{\bm{k}' \lambda'}} \Bigl( n_B(\hbar \omega_{\bm{k}'\lambda'}) + \frac{1}{2} \Bigr) \Bigr]
  \gamma^{\text{QHA}}_{\bm{k}' \lambda'}
\end{align}
Using the low-temperature and high-temperature form of the Bose distribution function $n_B$
\begin{align}
  n_B(\hbar \omega) + \frac{1}{2}
  \simeq
  \begin{cases}
    \frac{1}{2} + e^{-\beta\hbar \omega} &(\hbar \omega \gg k_{\mathrm{B}} T)\\
    \frac{k_{\mathrm{B}} T}{\hbar \omega} &(\hbar \omega \ll k_{\mathrm{B}} T),\\
  \end{cases}
\end{align}
we get 
\begin{align}
  \frac{\partial}{\partial \omega_{\bm{k}\lambda}}\Bigl[ \frac{1}{\omega_{\bm{k} \lambda}} \Bigl( n_B(\hbar \omega_{\bm{k}\lambda}) + \frac{1}{2} \Bigr) \Bigr]
  \simeq - \frac{1}{\omega_{\bm{k} \lambda}^2} \Bigl( n_B(\hbar \omega_{\bm{k}\lambda}) + \frac{1}{2} \Bigr) \times C,
\end{align}
where $C\simeq 1\sim2$ is a constant which is dependent on the mode and the temperature. Therefore, the estimation for the $\Delta_3 \gamma_{\bm{k}\lambda}$ is
\begin{align}
  \Delta_3 \gamma_{\bm{k}\lambda} \simeq \gamma^{\text{QHA}}_{\bm{k}\lambda} \times \Bigl( -C\frac{\braket{\hat{U}_4}}{\braket{\hat{U}_2}} \Bigr),
\end{align}
where $C$ is the average value over all the phonon modes.
To summarize the above discussion, the anharmonic correction to the Gr\"uneisen parameter is given by
\begin{align}
  \gamma_{\bm{k}\lambda} \simeq \gamma^{\text{QHA}}_{\bm{k}\lambda} \times \Bigl[ 1 -(2+C) \frac{\braket{\hat{U}_4}}{\braket{\hat{U}_2}}  +  \frac{P_4}{P_2} \Bigr]
  \label{Appendix_eq_gamma_correction}
\end{align}
The calculation of the anharmonic correction to the mode specific heat $c_{v,\bm{k}\lambda}$ is as shown in section \ref{subsec_perturbation_gamma_c_alpha}.
\begin{align}
  \Delta c_{v,\bm{k}\lambda} 
  &\simeq
  - c^{\text{QHA}}_{v,\bm{k}\lambda} \times \frac{1}{6} \frac{\braket{\hat{U}_4}}{\braket{\hat{U}_2}}
  \label{Appendix_eq_cvklambda_correction}
\end{align}

Before we calculate the correction to the thermal expansion coefficient, we consider the bulk modulus. We start with the QHA case. The QHA free energy is 
\begin{align}
  F_{\text{QHA}}
  &=
  \Phi_0 + \sum_{\bm{k}\lambda} \Bigl[ \frac{1}{2} \hbar \omega_{\bm{k}\lambda} + k_{\mathrm{B}} T \log (1- e^{-\beta \hbar \omega_{\bm{k}\lambda}}) \Bigr].
\end{align}
Thus the pressure and the bulk modulus can be derived as
\begin{align}
  P_{\text{QHA}} 
  &=
  - \frac{\partial F_{\text{QHA}}}{\partial V}
  \nonumber
  \\&=
  - \frac{\partial \Phi_0}{\partial V} - \sum_{\bm{k}\lambda} \hbar \frac{\partial \omega_{\bm{k}\lambda}}{\partial V} \Bigl( n_B(\hbar \omega_{\bm{k}\lambda}) + \frac{1}{2} \Bigr)
\end{align}
\begin{align}
  B_T^{\text{QHA}} 
  &=
  V \frac{\partial^2 \Phi_0}{\partial V^2} 
  +
  V \frac{\partial}{\partial V} \Bigl[ \sum_{\bm{k}\lambda} \hbar \frac{\partial \omega_{\bm{k}\lambda}}{\partial V} \Bigl( n_B(\hbar \omega_{\bm{k}\lambda}) + \frac{1}{2} \Bigr) \Bigr]
  \nonumber
  \\&=
  B_{T,0} + \Delta B_{T}^{\text{QHA}}
\end{align}
The correction to the bulk modulus from the vibrational free energy is known to be much smaller than the dominant term which comes from the electronic ground state energy, i.e. $|B_{T,0}| \gg |\Delta B_{T}^{\text{QHA}}|$~\cite{Wang_2010}.
Here, we calculate the SCP bulk modulus. By differentiating Eq. (\ref{Appendix_eq_SCP_pressure}), we get
\begin{align}
  B_T 
  =&
  V \Bigl[ \Bigl( \frac{\partial}{\partial V} \Bigr)_{T,\Omega} + \sum_{\bm{k}\lambda} \Bigl( \frac{\partial \Omega_{\bm{k}\lambda}}{\partial V} \Bigr)_T \frac{\partial}{\partial \Omega_{\bm{k}\lambda}} \Bigr]
  \nonumber
  \\&
  \Bigl[
  \frac{\partial \Phi_0}{\partial V}
  + \sum_{\bm{k}\lambda} 
  2\omega_{\bm{k}\lambda} \Bigl( \frac{\partial \omega_{\bm{k}\lambda}}{\partial V} \Bigr) 
  g(\Omega_{\bm{k}\lambda})
  \nonumber
  \\&+
  \sum_{n=2}^{\infty} \frac{1}{n!N^{n-1}} 
  \sum_{\bm{k}_1\lambda_1 \cdots \bm{k}_n\lambda_n} 
  \nonumber
  \\&\times
  \frac{\partial \widetilde{\Phi}(\bm{k}_1\lambda_1,-\bm{k}_1\lambda_1, \cdots, \bm{k}_n\lambda_n,-\bm{k}_n\lambda_n)}{\partial V}
  \nonumber
  \\&\times
  g(\Omega_{\bm{k}_1\lambda_1}) \cdots g(\Omega_{\bm{k}_n\lambda_n})
  \Bigr]
\end{align}
Thus,
\begin{align}
&
  \Delta B_T 
  = 
  B_T - B_T^{\text{QHA}}
  \nonumber
  \\=&
  V\Bigl( \frac{\partial}{\partial V} \Bigr)_T
  \sum_{\bm{k}\lambda} \Bigl[ \hbar \frac{\omega_{\bm{k}\lambda}}{\Omega_{\bm{k}\lambda}} \frac{\partial \omega_{\bm{k}\lambda}}{\partial V} \Bigl( n_B(\hbar \Omega_{\bm{k}\lambda}) + \frac{1}{2} \Bigr) 
  \nonumber
  \\&
  - \hbar \frac{\partial \omega_{\bm{k}\lambda}}{\partial V} \Bigl( n_B(\hbar \omega_{\bm{k}\lambda}) + \frac{1}{2} \Bigr) \Bigr]
  \nonumber
  \\&+
  V\Bigl( \frac{\partial}{\partial V} \Bigr)_T
  \Bigl[
  \sum_{n=2}^{\infty} \frac{1}{n!N^{n-1}} \sum_{\bm{k}_1 \lambda_1, \cdots, \bm{k}_{n}\lambda_{n}} 
  \nonumber
  \\&\times
  \frac{\partial \widetilde{\Phi}(\bm{k}_1\lambda_1,-\bm{k}_1\lambda_1, \cdots, \bm{k}_n\lambda_n,-\bm{k}_n\lambda_n)}{\partial V}
  \nonumber
  \\&\times
  g(\Omega_{\bm{k}_1\lambda_1}) \cdots g(\Omega_{\bm{k}_n\lambda_n})
  \Bigr]
  \nonumber
  \\&=
  \Delta B_{T,1} + \Delta B_{T,2}
  \label{Appendix_eq_Delta_BT_12}
\end{align}
The two terms in the RHS of Eq. (\ref{Appendix_eq_Delta_BT_12}) is estimated as follows.
\begin{align}
  \Delta B_{T,1} 
  \simeq&
  V\Bigl( \frac{\partial}{\partial V} \Bigr)_T
  \sum_{\bm{k}\lambda} \Bigl[ 
    \hbar \Bigl(\frac{\omega_{\bm{k}\lambda}}{\Omega_{\bm{k}\lambda}}-1\Bigr) \frac{\partial \omega_{\bm{k}\lambda}}{\partial V} \Bigl( n_B(\hbar \Omega_{\bm{k}\lambda}) + \frac{1}{2} \Bigr) 
    \nonumber
    \\&
    + 
    \hbar \frac{\partial \omega_{\bm{k}\lambda}}{\partial V} \Bigl( n_B(\hbar \Omega_{\bm{k}\lambda}) - n_B(\hbar \omega_{\bm{k}\lambda}) \Bigr) \Bigr].
\end{align}
Because it is possible to use the relation
\begin{align}
  n_B(\hbar \Omega_{\bm{k}\lambda}) - n_B(\hbar \omega_{\bm{k}\lambda}) 
  &\simeq
  - n_B(\hbar \omega_{\bm{k}\lambda}) \times \frac{\braket{\hat{U}_4}}{\braket{\hat{U}_2}}
\end{align}
for the phonon modes that contribute to the sum, we get
\begin{align}
  \Delta B_{T,1} 
  \simeq &
  -2 V \Bigl( \frac{\partial}{\partial V} \Bigr)_T \sum_{\bm{k}\lambda} 
  \Bigl[
    \frac{\braket{\hat{U}_4}}{\braket{\hat{U}_2}}
      \nonumber
  \\& \times
    \hbar \frac{\partial \omega_{\bm{k}\lambda}}{\partial V} \Bigl( n_B(\hbar \Omega_{\bm{k}\lambda}) - n_B(\hbar \omega_{\bm{k}\lambda}) \Bigr) 
  \Bigr]
  \nonumber
  \\\simeq &
  4\frac{\braket{\hat{U}_4}}{\braket{\hat{U}_2}} V \frac{\partial P_2}{\partial V} 
  \nonumber
  \\\simeq &
  -2 \frac{\braket{\hat{U}_4}}{\braket{\hat{U}_2}}\Delta B_{T}^{\text{QHA}},
\end{align}
where we assumed that $\frac{\partial}{\partial V}\frac{\braket{\hat{U}_4}}{\braket{\hat{U}_2}}$ is not very large. The other term in the RHS of Eq. (\ref{Appendix_eq_Delta_BT_12}) is estimated as
\begin{align}
  \Delta B_{T,2} 
  \simeq&
  V\Bigl( \frac{\partial}{\partial V} \Bigr)_T
  \Bigl[
   \frac{1}{2 N} 
   \sum_{\bm{k}_1 \lambda_1, \bm{k}_{2}\lambda_{2}} 
   \nonumber
  \\&\times
  \frac{\partial \widetilde{\Phi}(\bm{k}_1\lambda_1,-\bm{k}_1\lambda_1, \bm{k}_2\lambda_2,-\bm{k}_2\lambda_2)}{\partial V}
  \nonumber
  \\&\times
  g( \Omega_{\bm{k}_1\lambda_1}) g(\Omega_{\bm{k}_2\lambda_2})
  \Bigr]
  \nonumber
  \\\simeq&
  - V \Bigl( \frac{\partial P_4}{\partial V}\Bigr)_T.
\end{align}
Thus, 
\begin{align}
&
  \Delta B_T 
  \simeq
  \Delta B_T^{\text{QHA}} \times 
  \Bigl( - 2 \frac{\braket{\hat{U}_4}}{\braket{\hat{U}_2}} + \frac{1}{2}\frac{\partial P_4/\partial V}{\partial P_2/\partial V}\Bigr)
  \nonumber
  \\ &\ll
  | \Delta B_T^{\text{QHA}} | \ll | B_T |
\end{align}
Therefore, the anharmonic correction to the bulk modulus is much more minor than the correction to the other physical variables and thus is negligible in the calculation of the thermal expansion.

Therefore, as explained in section \ref{subsec_perturbation_gamma_c_alpha}, we get
\begin{equation}
    \alpha \simeq \alpha^{\text{QHA}}\times \Bigl( 1 - (\frac{13}{6}+C)\frac{\braket{\hat{U}_4}}{\braket{\hat{U}_2}} +\frac{P_4}{P_2} \Bigr).
    \label{Appendix_eq_alpha_perturbation}
\end{equation}

\subsection{The difference between the QHA volume and the SCP volume}
\label{Appendix_subsec_QHA_SCP_volume_diff}
Because the difference of the $V_{\text{QHA}}$ and $V_{\text{SCP}}$ can affect the calculation result through the change of IFCs, we consider the $V$-dependent change of the IFC $\Phi$.
\begin{eqnarray}
  \Phi(V_{\text{SCP}}) 
  &=&
  \Phi(V_0) + \int_{V_0}^{V_{\text{SCP}}} dV \frac{\partial \Phi}{\partial V}
  \\&=&
  \Phi(V_0) + \int_{V_0}^{V_{\text{QHA}}} dV \frac{\partial \Phi}{\partial V} + \int_{V_{\text{QHA}}}^{V_{\text{SCP}}} dV \frac{\partial \Phi}{\partial V},
\end{eqnarray}
where $V_0$ is the volume at the zero temperature, at which the difference of $V_{\text{QHA}}$ and $V_{\text{SCP}}$ can be neglected. 
When the thermal expansion coefficient is calculated with accuracy of 
$O({\braket{\hat{U}_4}}/{\braket{\hat{U}_2}})$, it is possible to estimate the IFC at $V_{\text{SCP}}$ as 
\begin{eqnarray}
  \Phi(V_{\text{SCP}}) 
  &\sim&
  \Phi(V_0) + \delta \Phi_{\text{QHA}} + \delta \Phi_{\text{QHA}} \times O\bigl(\frac{\braket{\hat{U}_4}}{\braket{\hat{U}_2}}\bigr)
\end{eqnarray}
where $\delta \Phi_{\text{QHA}} =  \int_{V_0}^{V_{\text{QHA}}} dV \frac{\partial \Phi}{\partial V}$. Since the harmonic phonon dispersion of materials in ambient condition does not drastically change with the temperature, we can conclude
\begin{equation}
  \Phi(V_0) \gg \delta \Phi_{\text{QHA}} \gg \delta \Phi_{\text{QHA}} \times O\bigl(\frac{\braket{\hat{U}_4}}{\braket{\hat{U}_2}}\bigr),
\end{equation}
at least for the harmonic IFCs. This shows that the order of the effect of the difference of $V_{\text{QHA}}$ and $V_{\text{SCP}}$ is smaller than $\Omega_{\bm{k}\lambda} - \omega_{\bm{k}\lambda}$. Therefore, the above discussion of the leading correction by the lattice anharmonicity, which implicitly assumes $V_{\text{QHA}} = V_{\text{SCP}}$, is justified.

\bibliography{apssamp}

\end{document}